\documentclass[aps,prc,reprint,showkeys,showpacs,nofootinbib,superscriptaddress,floatfix]{revtex4-1}

\usepackage[english]{babel}
\usepackage{amsmath,amssymb}
\usepackage{bm}
\usepackage{graphicx}
\usepackage{braket}
\usepackage[colorlinks=true,allcolors=blue]{hyperref}
\usepackage{microtype}

\providecommand{\dd}{\ensuremath{\mathrm{d}}}
\providecommand{\vv}[1]{\ensuremath{\mathbf{#1}}}

\begin{document}

\newcommand{\MeVfmcube}{$\mathrm{\:MeV\:fm}^3$}

\title{Global description of beta-minus decay in even-even nuclei with the
axially-deformed Skyrme finite amplitude method} \date{\today}
\author{M.\ T.\ Mustonen} \email{mika.mustonen@yale.edu}
\affiliation{Department of Physics and Astronomy, CB 3255, University of North Carolina, Chapel Hill, NC 27599-3255}
\affiliation{Center for Theoretical Physics, Sloane Physics Laboratory, Yale
University, New Haven, Connecticut 06502, USA}
\author{J.\ Engel} \email{engelj@physics.unc.edu}
\affiliation{Department of Physics and Astronomy, CB 3255, University of North
Carolina, Chapel Hill, NC 27599-3255}

\begin{abstract}
We use the finite amplitude method for computing charge-changing Skyrme-QRPA
transition strengths in axially-deformed nuclei together with a modern Skyrme
energy-density functional to fit several previously unconstrained parameters in
the charge-changing time-odd part of the functional. With the modified
functional we then calculate rates of beta-minus decay for all medium-mass and
heavy even-even nuclei between the valley of stability and the neutron drip
line.  We fit the Skyrme parameters to a limited set of beta-decay rates, a set
of Gamow-Teller resonance energies, and a set of spin-dipole resonance energies,
in both spherical and deformed nuclei.  Comparison to available experimental
beta-decay rates shows agreement at roughly the same level as in other global
QRPA calculations.  We estimate the uncertainty in our rates all the way to the
neutron drip line through a construction that extrapolates the errors of known
beta-decay rates in nuclei with intermediate $Q$ values to less stable isotopes
with higher $Q$ values.
\end{abstract}
\maketitle

\section{Introduction}

Beta-decay rates are an important ingredient in simulations of the
astrophysical $r$-process.  Because parts of the $r$-process path are still not
accessible to experiment, it is up to theoretical models to produce approximate
rates for many relevant neutron-rich isotopes.  Models could also help resolve
the issues raised by Ref.\ \cite{Mention2011}, which argued that the flux of
antineutrinos from nuclear reactors does not agree with the standard model.
Ref.\ \cite{Hayes2014} pointed out that the discrepancy could be due to an
overly simple treatment of first-forbidden beta decay in fission products. 

Several schemes/methods for calculating beta-decay rates across almost the
entire nuclear chart have been devised.  They include the proton-neutron
quasiparticle random phase approximation (pnQRPA) with a residual Gamow-Teller
interaction \cite{Homma1996}, gross theory \cite{Moller1997,Moller2003},
semi-gross theory \cite{Nakata1997}, the extended Thomas-Fermi plus Strutinsky
integral (ETFSI) method \cite{Borzov2000}, and artificial neural networks
\cite{Costiris2009}, and, very recently, the relativistic spherical pnQRPA
\cite{Marketin2015}.  Full beta-decay tables for neutron-rich isotopes have
been only published by M\"oller \emph{et al} \cite{Moller1997,Moller2003} and
Marketin \emph{et al} \cite{Marketin2015}.

Many other authors have applied more sophisticated and/or computationally
intensive methods to smaller sets of nuclei, focusing on some of those
important for the $r$ process.  Recent examples of such work include a deformed
pnQRPA computation with the Bonn-CD interaction of the decay of neutron-rich
isotopes with $Z = 36-43$ \cite{Fang2013}, of isotopes of Zr and Mo
\cite{Ni2014}, and of isotopes of Kr and Sr \cite{Ni2014a}; similar
calculations with Gogny interaction in the $N= 82,126,184$ isotonic chains
\cite{Martini2014}; relativistic pnQRPA \cite{Niu2013} for
$20 \le Z \le 50$; and relativistic pnQRPA for $N \approx 50, 82$
\cite{Niksic2005}. 

Computational barriers have thus far prevented the production of a beta-decay
table for the entire nuclear chart in a fully self-consistent Skyrme mean-field
approach that allows deformation.  Recently, however, we reported
\cite{Mustonen2014} an implementation of the charge-changing finite amplitude
method, which sidesteps the QRPA eigenvalue problem.  We obtain beta-decay
rates by directly computing the required sums and integrals over allowed final
states of the response to charge-changing perturbations.  We will soon make
available a code called \textsc{pnfam} that implements the method. 

We could proceed by choosing an existing density functional, interpreting it as
a density-dependent effective interaction, and calculating beta-decay rates. If
we were interested in, e.g., the effects of tensor terms discussed in Refs.\
\cite{Minato2013a} and \cite{De-Donno2014}, we could take them from
already-parametrized functionals.  Such a procedure would require some
parameter fitting because pairing interactions and strengths, especially those
associated with isoscalar pairing, are not usually specified alongside Skyrme
particle-hole effective interactions.  But that approach is still too limiting
because not all Skyrme functionals can be consistently represented as effective
interactions.  In particular, the time-even and time-odd parts of the
functional, which are related if the functional is the mean-field expectation
value of a Hamiltonian, need not be related in more general constructions.  

Our main goal here is to assess the ability of the Skyrme QRPA with deformation
to predict $\beta^-$ decay, and to use existing data (decay rates and resonance
energies) to constrain the isoscalar-pairing strength and the other time-odd
coupling constants, which in the general energy-density functional (EDF)
picture, are not fixed by fits to masses, precisely because they are
independent of the time-even functional.  In much of what follows, therefore,
we will not assume that the functional results from mean-field theory with an
interaction and so will have to fit a significant number of parameters.  After
presenting our methods and assessment, we display the (summarized) results of a
a full table of beta-minus rates, computed with the pnFAM, for even-even nuclei
in all medium-mass and heavy isotopic chains.  We use a simple but apparently
accurate model to quantify and extrapolate theoretical uncertainty. 

This article is organized as follows: Section~\ref{sec:theory} is a brief
overview of the theoretical background, and Section~\ref{sec:comp} details our
computational approach and parameter-fitting procedures. In
Section~\ref{sec:results}, we assess the quality of our results, comparing them
to earlier work and to experimental data where available.
Section~\ref{sec:conclusions} contains conclusions.

\section{Theoretical background}\label{sec:theory}

\subsection{Finite amplitude method}

The finite amplitude method (FAM), a formulation of the random-phase
approximation that speeds the computation of nuclear response functions, was
introduced in nuclear physics in Ref.~\cite{Nakatsukasa2007}.  It was later
generalized to the quasiparticle random phase approximation (QRPA) in
Ref.~\cite{Avogadro2011} and to the relativistic QRPA in
Ref.~\cite{Niksic2013}.  In Ref.~\cite{Mustonen2014} we applied the method to
charge-exchange transitions, in particular allowed and first-forbidden beta
decay.

The FAM solves equations for the amplitude of the linear response to a small
but finite perturbation.  As a result, the method does not directly yield the
poles and residues of the response, which are the central objects in the matrix
version of the QRPA.  But if the goal of the computation is to get transition
strength functions in a large model space the FAM can yield results in orders
of magnitude less CPU time than matrix QRPA.

The FAM offers another advantage for beta decay: the weighted sums or integrals
of transition strength can be expressed as contour integrals.  This fact was
first exploited by Hinohara et al.~\cite{Hinohara2015}, who evaluated the
response at a relatively small number of complex frequencies to compute sum
rules.  In Ref.~\cite{Mustonen2014} we used the idea to evaluate the more
complicated beta-decay phase-space-weighted integrals, which are not analytic
and contain interference terms between first-forbidden operators.  With the FAM
we can thus use typical supercomputer resources to calculate many observables
in a large number of nuclei.  We are able to extend systematic Skyrme parameter
fitting from mean-field calculations to deformed QRPA calculations, at least in
a preliminary way.

\subsection{Model parameters and fitting targets}

Our starting point in the particle-hole channel is a generic Skyrme EDF:
\begin{equation}\label{eq:skyrme-edf}
\mathcal E = \sum_{t=0,1} \sum_{t_3 = -t}^{+t}
\int \dd\vv r \, \boldsymbol{\left(}\mathcal{H}_{tt_3}^\mathrm{even}(\vv r) +
\mathcal{H}_{tt_3}^\mathrm{odd}(\vv r) \boldsymbol{\right)} \,.
\end{equation}
Here $\mathcal{H}_{tt_3}^\mathrm{even}$ contains products of time-even local
densities only, with coefficients fixed by fitting to masses and perhaps a few
other quantities, and $\mathcal{H}_{tt_3}^\mathrm{odd}$ is given by 
\begin{equation}
\label{eq:h_odd}
\begin{aligned}
\mathcal{H}_{tt_3}^\mathrm{odd}(\vv r) &\equiv
C_t^s[\rho_{00}]\vv s_{tt_3}^2 +
C_t^{\Delta s}\vv s_{tt_3} \cdot \bm\nabla^2 \vv s_{tt_3} +
C_t^{j} \vv j_{tt_3}^2 \\ & + 
C_t^T \vv s_{tt_3} \cdot \vv T_{tt_3} + 
C_t^{s\nabla j} \vv s_{tt_3} \cdot \bm\nabla \times \vv j_{tt_3} \\ & +
C_t^F \vv s_{tt_3} \cdot \vv F_{tt_3} + 
C_t^{\nabla s} \left(\bm\nabla \cdot \vv s_{tt_3}\right)^2 \,,
\end{aligned}
\end{equation}
with the spin density $\vv s_{tt_3}$, the current density $\vv j_{tt_3}$, the
spin-kinetic density $\vv T_{tt_3}$, and the tensor-kinetic density $\vv F_{tt_3}$
defined e.g.\ in Ref.\ \cite{Perlinska2004}.  If one requires the functional to
be the mean-field expectation value of a Skyrme interaction, EDF coupling
constants are completely determined by the (fewer) parameters that specify the
interaction, as discussed in Refs.\ \cite{Perlinska2004,Bender2002} and
mentioned above.  In this work, we adopt the view that the effective
``interaction'' comes from the energy-density functional rather than the other
way around.  Consequently we are free to fit all the time-odd coupling
constants without spoiling the mass fits generated the time-even couplings.  We
do, however, adopt the values obtained from the Skyrme parametrization as our
starting point for the fits, unless we note otherwise.

The subset $\{ C^s_1, C^T_1, C^F_1 \}$ of time-odd coupling constants maps
directly to the parameters of Landau-Migdal interaction for infinite
homogeneous nuclear matter with tensor terms \cite{Backman1979a}.  In that
sense, these couplings are intimately related to the bulk properties of the
nuclear matter.  The constant $C^F_1$ is purely tensor in character, and it
determines the tensor term in the Landau-Migdal interaction.  The spin-density
coupling constant $C^s_1$ strongly affects the Gamow-Teller strength
distribution and can be fit to the locations of Gamow-Teller resonances
\cite{Bender2002}.  The last term $C^T_1$ maps to a linear combination of the
Landau parameters of both the tensor term and a term that depends on the
scattering angle of the Laundau-Migdal quasiparticles. 

Two other parameters have a large effect in the QRPA: the strengths of the
residual particle-particle (or pairing) interaction between protons and
neutrons,
\begin{equation}\label{eq:vpp}
V_\mathrm{pp} = \left( V_0 \hat\Pi_{T=0} + V_1 \hat\Pi_{T=1} \right) \left( 1 - \alpha \frac{\rho_{00}(\vv r)}{\rho_c}  \right) \delta(\vv r),
\end{equation}
where $\rho_c = 0.16$~fm$^{-3}$ is the saturation density of nuclear matter and
$\alpha \in [0,1]$ controls the pairing density-dependence (throughout this work
we use mixed pairing, i.e.\ $\alpha = 0.5$). The isovector ($T=1$)
proton-neutron pairing mainly affects Fermi beta decay, which plays almost no
role in heavy nuclei; we simply set its strength $V_1$ to be the average of the
neutron-neutron and proton-proton pairing strengths (both fixed in the HFB part
of the calculation).  The isoscalar ($T=0$) pairing is a different story; it has
a strong effect on Gamow-Teller decay, and determining a reasonable value for
its strength is a common issue in single and double beta decay computations.
The HFB mean field is independent of the $T=0$ pairing term as long as explicit
proton-neutron mixing is neglected.  We are thus free to fit $V_0$ in the QRPA. 

We include three types of observables in our fitting procedure: Gamow-Teller
resonance energies, spin-dipole resonance energies, and total beta decay rates,
all in both spherical and deformed isotopes.  Although we pick several sets of
target nuclei for fitting the decay half-lives, each set includes a large range
of mass values so that our fits can be global.  To assess the success of the
fits and to compare them to earlier work in very different models, we also
compute the following metrics for beta-decay tables, as laid out e.g.\ in
\cite{Moller1997,Moller2003}: the residual of each computed $\lg t$ value (where
$t$ is the half-life and the logarithm is 10-based), 
\begin{equation}\label{eq:r1def}
  r = \lg \left( \frac{t_\textrm{th.}}{t_\textrm{exp.}} \right),
\end{equation}
the average of the residuals,
\begin{equation}
  M_{r} = \frac{1}{n} \sum_{i=1}^n r_i,
\end{equation}
(where we've used an index $i$ to indicate that there is an $r$ for every
nucleus) and the standard deviation of the residuals around the average,
\begin{equation}
  \sigma_{r} = \sqrt{\frac{1}{n} \sum_{i=1}^n \bigl[ (r_i - M_{r} \bigr]^2 }.
\end{equation}
Ref.\ \cite{Costiris2009} contains an excellent compilation of these quantities.

\subsection{Uncertainty analysis}

As theoretical approaches grow more sophisticated, the analysis of theoretical
uncertainty is growing in importance.  Here we attempt to provide reasonable
estimates for the uncertainty in our predicted half-lives, particularly in
isotopes that are too short-lived to allow measurement.

The standard prescription for assigning a theoretical (statistical) uncertainty
$\Delta \mathcal{O}$ to a computed observable $\mathcal{O}$ is
\cite{Dobaczewski2014}
\begin{equation}\label{eq:uncertainty}
  \Delta \mathcal{O} = \sqrt{\sum_{ab} \frac{\partial \mathcal{O}}{\partial x_a} \bigg|_{x=x_0} C_{ab} \frac{\partial \mathcal{O}}{\partial x_b}\bigg|_{x=x_0}},
\end{equation}
where $C$ is the covariance matrix
\begin{equation}
  C = (J^T J)^{-1}
\end{equation}
and $x = (x_1, \ldots, x_{N_x})$ are the $N_x$ parameters of the model.  The
partial derivatives of all the observables $\{\mathcal{O}_a\}$ with respect to
all the parameters evaluated at the result of the fit $x_0$ form the Jacobian
$J$:
\begin{equation}\label{eq:jacobian}
  J_{ab} = \frac{\partial \mathcal{O}_a}{\partial x_b}\bigg|_{x=x_0}.
\end{equation}
When not analytically accessible, the needed partial derivatives can be
estimated through finite central differences.  (In general, the theoretical
uncertainty related to the Jacobian must be supplemented by numerical and
experimental uncertainties.  We have assumed that the theoretical uncertainty is
much larger than the other two, which we therefore neglect.)

To use Eq.\ \eqref{eq:uncertainty} to assign an uncertainty to every beta-decay
rate in our table, we would need to evaluate the Jacobian in Eq.\
\eqref{eq:jacobian} for (the logarithms of) all the half-lives $t$ in our table.
Unfortunately, the required $2N_x$ full decay-table computations are still not
possible in a reasonable amount of computer time, even with the efficiency of
the FAM.  We therefore attempt to gauge the uncertainties and their $Q$
dependence in a more na\"ive way.  We construct a simple few-parameter
\emph{model for the uncertainties} that we can then fit to the observed
differences between our numerical results and known experimental values.  The
resulting approach is agnostic about how the decay is actually calculated; it
treats the nuclear model as a black box that produces predictions for $Q$ values
and beta-decay rates.  It does, however, make several assumptions about both the
calculated and experimental strength distributions that are only approximately
correct.

The first assumption is that the final states that contribute significantly to
a decay rate lie not too far from the ground state in a relatively small window
of excitation energy, so that we can reasonably approximate them, again either
in our calculation or in the real world, by one effective state with an
effective $Q$ value $q_\mathrm{eff}$ (that is not too different from the
ground-state $Q$ value):
\begin{equation}
  q_\mathrm{eff} = \frac{\sum_k C_k f_k(q_k + 1, Z_f) \, q_k}{\sum_k C_k
  f_k(q_k + 1, Z_f)} \,.
\end{equation}
Here $C_k$ is the standard integrated shape function for the transition to the
final state $k$ --- for an allowed state decay this is simply the transition
strength, and for a non-unique forbidden decay it encapsulates several
terms --- and $f_k$ is the usual allowed phase-space
integral.  The quantity $q_k$ is the $Q$ value of the decay to the state $k$
divided by $m_e c^2$, and $Z_f$ is the charge of the daughter nucleus.

With these definitions, we can proceed to define an effective shape factor
$C_\mathrm{eff}$ as
\begin{equation}\label{eq:halflife}
  t = \frac{\kappa}{\sum_k C_k f(q_k + 1, Z_f)}
  = \frac{\kappa}{C_\mathrm{eff} f(q_\mathrm{eff} + 1, Z_f)}.
\end{equation}
The $C_k$ depend on the $q_k$ for forbidden transitions, but the dependence is
weak compared to that of the corresponding phase-space integral, and so we
neglect it in our effective shape factor.

All these definitions can be made independently for the experimental strength
distribution and the theoretical one.  The quantity we wish to understand is the
ratio $r$ of the theoretical and experimental lifetimes:
\begin{equation}
\label{eq:lifetime-ratio}
\begin{split}
  r =& \lg \frac{t_\textrm{th}}{t_\textrm{exp}} \\
  =& \lg \frac{C_\textrm{eff}^\mathrm{exp}}{C_\textrm{eff}^\textrm{th}} \\
  &+ \lg f(q_\textrm{eff}^\textrm{exp}+1,Z_f) - \lg
  f(q_\textrm{eff}^\textrm{th} + 1,Z_f) \,,
\end{split}
\end{equation}
where the meanings of $q_\textrm{eff}^\textrm{exp}$ and
$q_\textrm{eff}^\textrm{th}$ and the corresponding quantities
$C_\textrm{eff}^\mathrm{exp}$ and $C_\textrm{eff}^\textrm{th}$ should be clear.
We omit the nucleus index $i$ in them for brevity.  Because the phase space
grows quickly with decay energy, the effective $Q$ values will usually be close
to ground-state-to-ground-state $Q$ value.  We therefore expand both logarithms
in Eq.\ \eqref{eq:lifetime-ratio} of the phase-space factors about the
theoretical ground-state-to-ground-state $Q$ value $q_\textrm{g.s.}^\textrm{th}$
to first order in $q$: 
\begin{equation}\begin{split}
  \lg f(q+1, Z_f) \approx & \lg f(q^\textrm{th}_\textrm{g.s.} + 1, Z_f) \\
  & + \frac{d \lg f(q+1,Z_f)}{dq} \biggr|_{q=q^\textrm{th}_\textrm{g.s.}} \hspace{-.5cm} \cdot (q - q^\textrm{th}_\textrm{g.s.}) \\
  =& \lg f(q^\textrm{th}_\textrm{g.s.} + 1, Z_f) \\
  & + \frac{f'(q^\textrm{th}_\textrm{g.s.} + 1,
  Z_f)}{f(q^\textrm{th}_\textrm{g.s.} + 1, Z_f)} \cdot \frac{q -
  q^\textrm{th}_\textrm{g.s.}}{\ln 10}.  
  \end{split}\end{equation}
This approximation is best when the $Q$ value is high, because the curvature of
$\lg f(q+1,Z_f)$ is small at high $q$.  For $Q$ values lower than about 2--3~MeV
the first-order approximation is poor, so we exclude such data points from our
analysis.  

Replacing the two logarithms in Eq.\ \eqref{eq:lifetime-ratio} by the
first-order expressions, we find that several terms cancel in the difference, so
that 
\begin{equation}\begin{split}
   r \approx & \lg \frac{C_\textrm{eff}^\textrm{exp}}
   {C_\textrm{eff}^\textrm{th}} \\
  &+ \frac{f'(q^\textrm{th}_\textrm{g.s.} + 1,
  Z_f)}{f(q^\textrm{th}_\textrm{g.s.} + 1, Z_f)} \cdot
  \frac{q_\textrm{eff}^\mathrm{exp} - q_\textrm{eff}^\textrm{th}}{\ln 10}
  \,.
\end{split}\end{equation}

We now make another set of assumptions, this time about the distribution of the
errors in the theoretical values: First, we assume that the relative error in
the effective shape factor is normally distributed with a slight systematic
component.  That is, we assume that for each nucleus, we have
\begin{equation}
\lg \frac{C^\textrm{exp}_\textrm{eff}}{C^\textrm{th}_\textrm{eff}} = c + \delta
c,
\end{equation}
where $c$ is a constant, independent of the nucleus in which the decay occurs,
that captures the systematic error and the set of all nucleus-dependent $\delta
c$'s is normally distributed with standard deviation $\Delta c$, which is a
measure of statistical error that is independent of decay energy.  Similarly, we
assume that the errors in the theoretical effective $Q$ value follow a normal
distribution, so that 
\begin{equation}
  \frac{q_\textrm{eff}^\textrm{exp} - q_\textrm{eff}^\textrm{th}}{\ln 10} = q +
  \delta q,
\end{equation}
where $q$ is now a nucleus-independent systematic error in the effective $Q$
value and the set of $\delta q$'s (again, one for each nucleus) is normally
distributed around zero with the standard deviation $\Delta q$ that represents
statistical error, again independent of $Q$. (Both $q$ and $\delta q$ are
expressed in units of $m_e c^2$, and the factor $1/\ln 10$ is absorbed into $q$
and $\delta q$ for convenience.)  Finally, we assume that the errors $\delta q$
and $\delta c$ are independent.

With these assumptions, we then have
\begin{equation}\label{eq:rmodel}
r = c + \frac{f'(q^\textrm{th}_\textrm{g.s.} + 1, Z_f)}
{f(q^\textrm{th}_\textrm{g.s.} + 1, Z_f)} \cdot q + \delta r,
\end{equation}
where $\delta r$ is a random error, the set of which must be normally
distributed at each $q^\textrm{g.s.}_\textrm{th}$ with width
\begin{equation} 
\label{eq:stdev}
\Delta r (q^\textrm{th}_\textrm{g.s.})^2  = {\Delta c^2 + \biggl(
\frac{f'(q^\textrm{th}_\textrm{g.s.} + 1,
Z_f)}{f(q^\textrm{th}_\textrm{g.s.} + 1, Z_f)} \biggr)^2 \Delta q^2}.
\end{equation}
We can now use Eqs.\ \eqref{eq:rmodel} and \eqref{eq:stdev} to determine the
values of $c$, $q$, $\Delta c$, and $\Delta q$.  We obtain the first two
through a fit to Eq.\ \eqref{eq:rmodel}, which expresses $r_i$ as a function of
$f'/f$ (and thus of $q_\textrm{g.s.}^\textrm{th}$), with the set of ratios $r_i$
given by our calculations (and experiment) and $\delta r_i$ set to zero.
Finally, we insert the fit values of $c$ and $q$ into the square of Eq.\
\eqref{eq:rmodel}, which then expresses $\delta r_i^2 $ as a function of
$(f'/f)^2$, and determine the values of $\Delta c$ and $\Delta q$ by requiring
that the line for $\Delta r(q_\textrm{g.s.}^\textrm{th})^2$ as a function of
$(f'/f)^2$ that expresses Eq. \eqref{eq:stdev} goes through the middle of the
data, with as many points below $\Delta r^2$ as above it.  A second least
squares fit of $\Delta c$ and $\Delta q$ in the right hand side of Eq.\
\eqref{eq:stdev} to the set of $\delta r_i^2$'s accomplishes the task nicely.
The explicitly $Q$-dependent $\Delta r$ can then be extrapolated to large $Q$
values where there are no data.

For both fits, in practice, we only include data points with $Q \ge 3$~MeV and
exclude any data points with $|r_i| > 4$ as obvious outliers.  We also adopt the
Primakoff-Rosen approximation \cite{Primakoff1959} for the phase-space integral
in the uncertainty model, making the expression $f'/f$ a simple ratio of two
polynomials, independent of $Z_f$:
\begin{equation}
  \frac{f'(x, Z_f)}{f(x, Z_f)}
  \approx \frac{5x^4 - 20x + 15}{x^5 - 10x^2 + 15x - 6}.
\end{equation}
Once the four parameters of the uncertainty model have been determined, the
uncertainties of the theoretical predictions are obtained the inverse relation
of \eqref{eq:r1def}
\begin{equation}
  t_\mathrm{exp} = t_\mathrm{th} \cdot 10^{r}
\end{equation}
with
\begin{equation}
  r = r(q^\textrm{th}_\textrm{g.s.}) \pm \Delta r(q^\textrm{th}_\textrm{g.s.}).
\end{equation}

\section{Skyrme functional and computational method}\label{sec:comp}

\begin{figure}[tb]
  \includegraphics[width=\columnwidth]{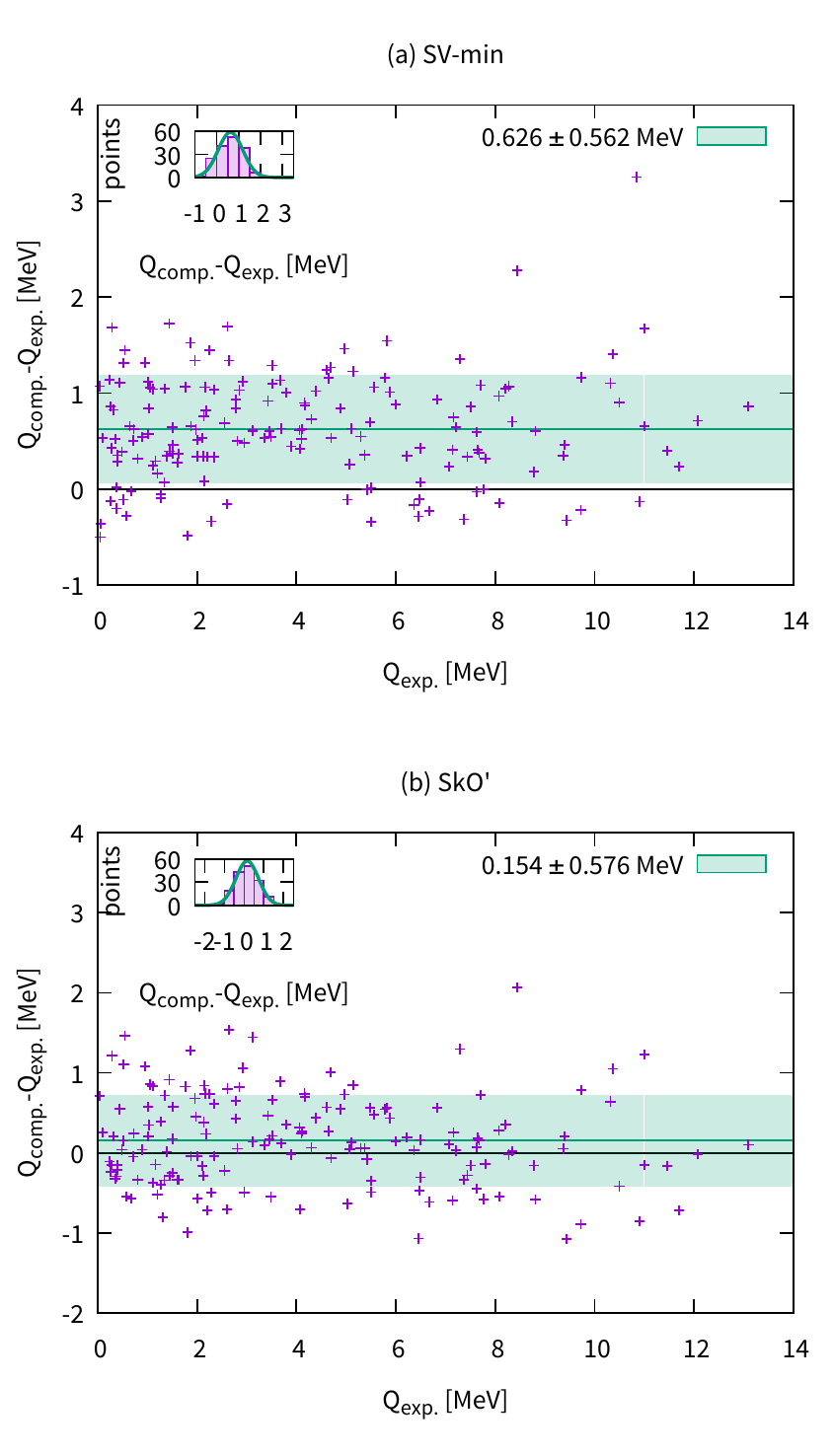}
  \caption{ The differences between computed and experimental $Q$ values, with
  SkO' (top) and with SV-min (bottom).  The insets show the distribution of
  differences.  The errors in our computed $Q$ values follow a normal
  distribution with an average of $0.154$~MeV and a standard deviation of
  $0.576$~MeV, with no noticeable bias when moving to higher $Q$ values.
  }\label{fig:qvalues}
\end{figure}

\begin{table*}[t]
  \begin{ruledtabular}
  \begin{tabular}{lccl}
    set & GT resonances & SD resonances & beta-decay half-lives \\
    \hline
    A & $^{208}$Pb, $^{112}$Sn, $^{76}$Ge, $^{130}$Te, $^{90}$Zr, $^{48}$Ca & none & $^{48}$Ar, $^{60}$Cr, $^{72}$Ni, $^{82}$Zn, $^{92}$Kr, $^{102}$Sr, $^{114}$Ru, $^{126}$Cd, $^{134}$Sn, $^{148}$Ba \\
    B & same as A & none & $^{52}$Ti, $^{74}$Zn, $^{92}$Sr, $^{114}$Pd, $^{134}$Te, $^{156}$Sm, $^{180}$Yb, $^{200}$Pt, $^{226}$Rn, $^{242}$U \\
    C & same as A & none & $^{52}$Ti, $^{72}$Ni, $^{92}$Sr, $^{114}$Ru, $^{134}$Te,$^{156}$Nd, $^{180}$Yb, $^{204}$Pt, $^{226}$Rn, $^{242}$U \\
    D & those of A and $^{150}$Nd & none & $^{58}$Ti, $^{78}$Zn, $^{98}$Kr, $^{126}$Cd, $^{152}$Ce,$^{166}$Gd, $^{204}$Pt \\
    E & same as D & $^{90}$Zr, $^{208}$Pb & $^{58}$Ti, $^{78}$Zn, $^{98}$Kr, $^{126}$Cd, $^{152}$Ce, $^{166}$Gd, $^{226}$Rn \\
  \end{tabular}
  \end{ruledtabular}
  \caption{The sets of fitting targets used in this work. The beta-decay
  half-lives in set A range from $0.069$~s ($^{102}$Sr) to $1.84$~s ($^{92}$Kr),
  in set B from $95.6$~s ($^{74}$Zn) to $45360$~s ($^{200}$Pt), in set C from
  $0.54$~s ($^{114}$Ru) to $9399.6$~s ($^{92}$Sr), and in set E from $0.046$~s
  ($^{98}$Kr) to $444$~s ($^{226}$Rn). The nuclei selected for fitting the
  beta-decay half-lives in sets D and E all exhibit an excitation spectrum
  clearly associated with either a spherical or a well-deformed shape; set E
  only consists of open-shell nuclei. The experimental Gamow-Teller resonance
  energies are from Refs.\
  \cite{Akimune1995,Anderson1985,Madey1989,Pham1995,Wakasa1997}, the spin-dipole
  resonance energies from Refs.\ \cite{Yako2006,Wakasa2012}, and the half-lifes
  from Ref.~\cite{ensdf2014apr}.  \label{tab:fit-sets}}
\end{table*}

The first step in our computational procedure is the construction of ground
states in the doubly-even mother nucleus with \textsc{hfbtho}, a
well-established HFB solver working in a (transformed) harmonic-oscillator basis
\cite{Stoitsov2013}.  We cut off the single-particle space at $60$ MeV to avoid
divergences from our zero-range pairing.  For each nucleus, we search for a
prolate, an oblate, and a spherical solution, and take the most bound of these
to be the ground state.

Because the beta-decay rates are very sensitive to the $Q$ value of the decay,
we look for a modern Skyrme functional that reproduces $Q$ values well.  Because
we don't explicitly treat odd nuclei, we use the prescription of Ref.\
\cite{Engel99} to approximate the $Q$ value; we have checked the prescription
against odd-A calculations in the equal filling approximation, and the two
procedures generally agree to within about 0.5~MeV. Of the several functionals
we examine, SkO' \cite{Reinhard1999} (with the strengths of proton-proton and
neutron-neutron pairing fit to the experimental pairing gaps of ten isotopes
picked in a wide mass range $50 \le A \le 230$) does the best job with $Q$
values, producing errors for ground-state-to-ground-state $Q$ values that are
normally distributed, with an average systematic error of $0.154$ MeV and
statistical error of $0.576$ MeV.  Figure \ref{fig:qvalues} compares the $Q$
values produced by SkO' with those of the next best functional, SV-min.

To compute beta-decay rates and resonance energies we use the code
\textsc{pnfam}, an implementation of the charge-changing finite amplitude method
presented in \cite{Mustonen2014}.  Built to work together with \textsc{hfbtho},
\textsc{pnfam} allows us to compute properties of axial deformed nuclei,
including both allowed and first-forbidden beta decay. 

We obtain our most robust fits by fixing all but two of the time-odd coupling
constants of the functional at values implied by its interpretation as an
interaction. (The mapping of the Skyrme coupling constants to the energy-density
functional coupling constants is explicitly discussed in
Refs.~\cite{Perlinska2004,Bender2002}.) The exceptions are $C^s_1$ and
$C^{\Delta s}_1$. We set the latter to zero to avoid known finite-size
instabilities \cite{Schunck2010} which, in the case of \textsc{hfbtho} and
\textsc{pnfam}, manifest themselves as divergences in the iterative solution.
That leaves $C^s_1$, which, along with the isoscalar pairing strength $V_0$, we
fit to a set of Gamow-Teller resonance energies, spin-dipole resonance energies,
and beta-decay rates selected from a wide mass range with no particular region
favored.  We use the code \textsc{pounders}, based on a derivative-free
algorithm \cite{Munson2012} designed for optimizing computer-time-consuming
penalty functions, to efficiently minimize the weighted the sum of the least
squares, simultaneously fitting both parameters.  We take $C^s_1$ to be
independent of the density and $V_0$ to have the same density dependence as the
proton and neutron pairing.  The axial-vector coupling constant $g_\textrm{A}$
is known to be quenched in nuclei, but the source and magnitude of the quenching
is an open problem; a variety of very different values for an effective
$g_\textrm{A}$ have been used.  For lack of a better prescription, we use the
commonly adopted quenched value $g_\textrm{A} = 1.0$ in the Gamow-Teller
channel, while applying no quenching in in the first-forbidden channels.  We
weight the three types of observables --- two kinds of energies and a rate ---
in the least-squares fit so as to approximately normalize the total penalty
function $\chi^2$ to the number of degrees of freedom, following the
recommendation in Ref.\ \cite{Dobaczewski2014}.  We assume that the theoretical
error dominates the experimental error, and thus assign equal weight to
observable of the same type.  We select these weights based on
how well the different observables are reproduced in an initial test fit, so
that each type of observable is approximately equally weighted in the actual
fit. Typical fits then take 10---20 thousand CPU hours, and we use XSEDE
supercomputers \cite{Towns2014} to carry them out.

Following the fit, we proceed to compute the beta-decay rates of all even-even
neutron-rich nuclei with $28 \le Z \le 110$, $A \ge 50$, all the way to the
neutron drip line, omitting just a few very stable isotopes for which the $Q$
value is negative in our HFB calculations.

\begin{figure*}
  \includegraphics[width=\textwidth]{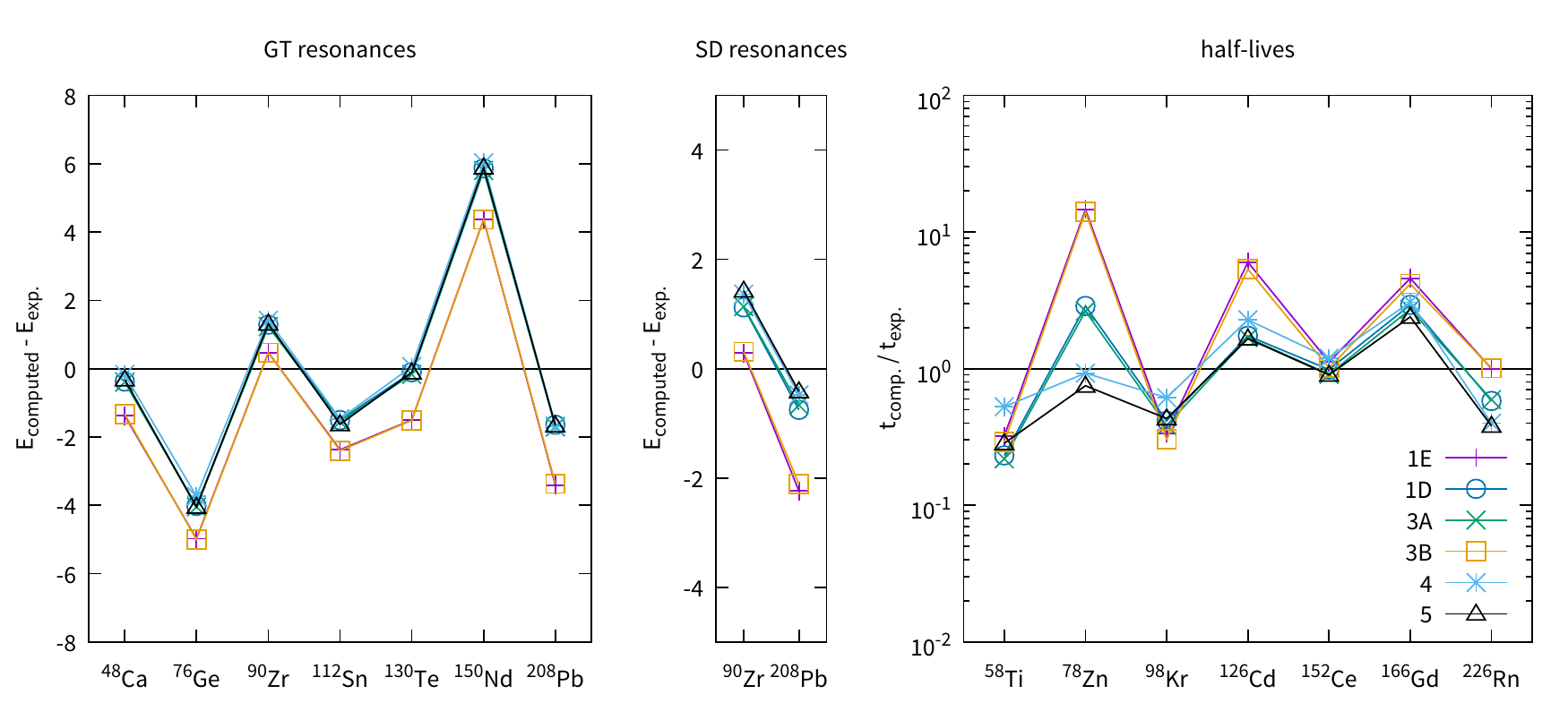}
  \caption{Reproduction of target data in those of our fits that use the target
  set E. The four-parameter fits 3A and 3B yield almost the same results as the
  two-parameter fits 1D and 1E, with only a small decrease in the penalty
  function.
  }\label{fig:fitting-targets}
\end{figure*}

\section{Results and discussion}\label{sec:results}

\subsection{Fit results}

\begin{figure}
  \includegraphics[width=\columnwidth]{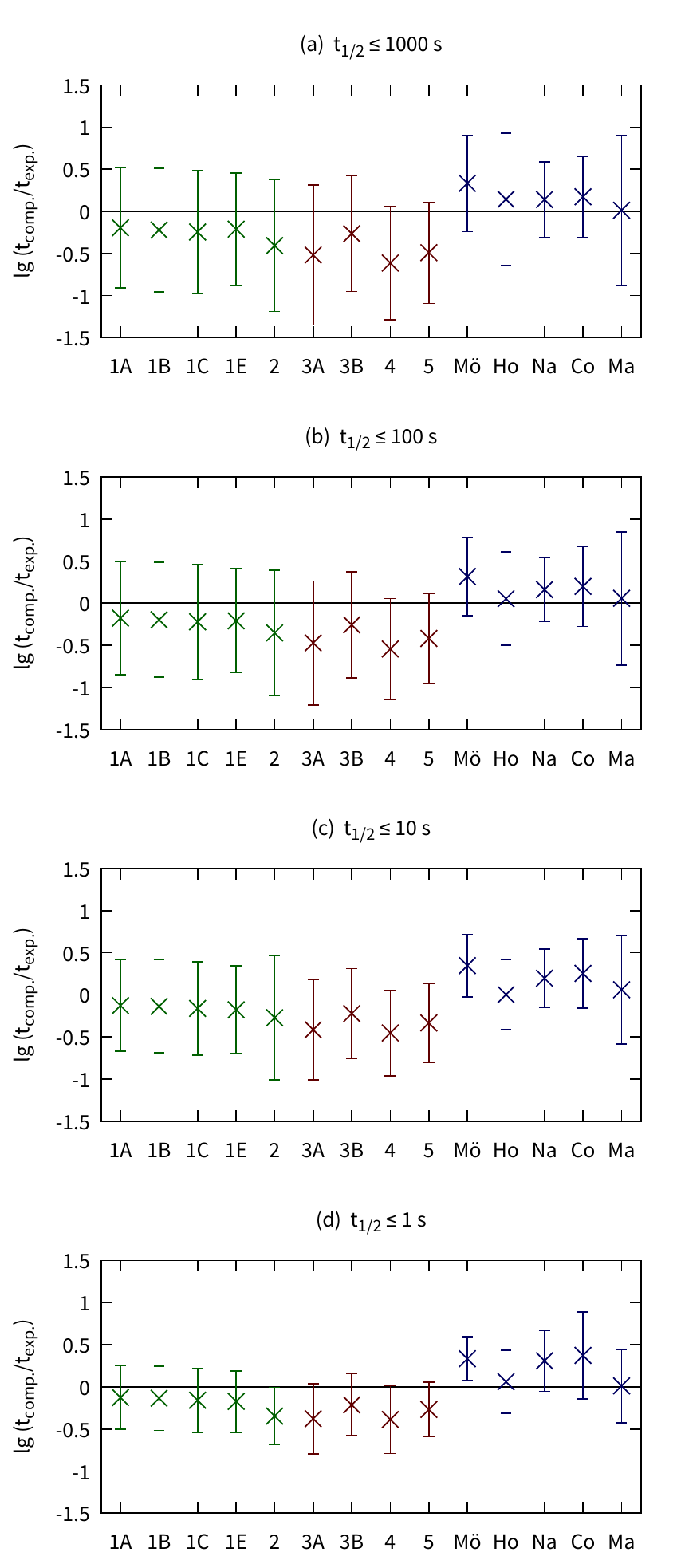}
  \caption{ Comparison of the mean and standard deviation of the $\lg t$ values
  in our fits with those of previous work.  The labels for our fits correspond
  to those of Table \ref{tab:fit-sets}.  The results from prior work are
  contained in \cite{Moller2003} (M\"o), \cite{Homma1996} (Ho),
  \cite{Nakata1997} (Na), \cite{Costiris2009} (Co), and \cite{Marketin2015}
  (Ma).  Only even-even isotopes are considered.
  }\label{fig:comparison-fits}
\end{figure}

To assess how sensitive our fit is to the set of target beta-decay rates and
resonance energies, we repeat the process with four sets of rates, summarized in
Table~\ref{tab:fit-sets}.  Each of these sets spans a large range of masses.
Set A contains beta-decay isotopes with relatively short half-lives only, set B
relatively long half-lives only, and set C a wide range of half-lives. (Short
half-lives should be less sensitive to details in nuclear structure).  Set D
contains only nuclei that are known to be rather rigid, with an excitation
spectrum characteristic of a spherical or a well-deformed nucleus.  (The QRPA,
which is based on a single mean field, should work best in rigid nuclei).  In
set E, we include only open-shell rigid nuclei (for which isoscalar pairing
should be most effective), swapping out $^{204}$Pt for $^{226}$Rn, and including
two spin-dipole resonances.  Figure \ref{fig:fitting-targets} shows the quality
of the fits for set E both with computed (1D) and experimental (1E) $Q$ values
(see Table \ref{tab:fits} for definitions of the number-letter combinations).
The two procedures yield very similar results.  The comparison in Fig.\
\ref{fig:comparison-fits}, which shows the results of the two-parameter fits
(along with those of more-parameter fits discussed shortly) when the resulting
functionals are applied to the set of all measured even-even $\beta^-$-decay
half-lives, shows that all these two-parameter fits (1A through 1E) yield the
same level of predictive accuracy. 

Can we do better by including some of the other time-odd coupling constants in
our fit?  To find out, we refit the four-parameter set $\{ V_0, C^s_1, C^T_1,
C^F_1\}$, which determines the Landau parameters $\{ g_0', g_1', h_0' \}$ and
thus allows us to incorporate infinite-nuclear-matter stability conditions
\cite{Backman1979a} as constraints.  The parameter $C^F_1$ introduces a time-odd
tensor term that is not present in the two-parameter fits.  The result, however,
improves the description of the fitting targets only marginally, as the points
labeled 3A and 3B in Fig.\ \ref{fig:fitting-targets} show, and actually worsens
the agreement with half-life measurements overall (as Fig.\
\ref{fig:comparison-fits} shows).  The situation gets even worse when we use the
results of this fit as a starting point to fit three more time-odd coupling
constants, $\{ C^j_1, C^{\nabla j}, C^{\nabla s} \}$ (fit 4). Then the
beta-decay rates to which we fit are reproduced better, but the agreement with
all measured rates deteriorates. 

Improvement in the fitting targets accompanying a deterioration in overall
agreement with data is a symptom of overfitting.  To better understand why this
happens, we evaluate the Jacobian matrix \eqref{eq:jacobian} at the parameter
values produced by the two-parameter fit 1E.  The Jacobian appears in
Table~\ref{tab:jacobian}, with the values of the coupling constants in natural
units following the prescription of Ref.\ \cite{Kortelainen2010} and the natural
scale of isoscalar pairing taken to be the strength of isovector pairing.  A
clear column structure appears in both the resonance energies and the
half-lives, signaling that the members of each individual set move largely in
unison when the parameters are varied.  Thus there are essentially just two
meaningful degrees of freedom that we can expect to fix with this experimental
data: $V_0$ and $C^s_1$.  

To see this in more detail, we carry out a singular value decomposition of the
Jacobian.  The largest singular value, $122.53$, corresponds to a vector
pointing nearly in the direction of $C^s_1$ in parameter space, and the second
largest, $10.85$, to a vector pointing nearly in the direction of $V_0$.  The
third largest value, $1.648$, is almost two orders of magnitude smaller than the
largest, and corresponds mostly to $C^T_1$, with many other directions mixed in.
The charge-changing data we have available --- Gamow-Teller and spin-dipole
resonances, and half-lives --- are not enough to reasonably constrain more than
the parameters $V_0$ and $C^s_1$ in our initial fit. 

Figure \ref{fig:comparison-fits}, besides containing the results of our fits,
contains results from other work: Refs.\
\cite{Homma1996,Nakata1997,Costiris2009} and \cite{Marketin2015}.  Of all the
these computations, the one by Homma et al.\ \cite{Homma1996} seems to best
reproduce the known $\beta^-$ half-lives, even though it neglects non-unique
first-forbidden decay and uses simple separable interactions.  As
Figure~\ref{fig:multipole-decomp} shows, in our computation the non-unique $1^-$
contribution is quite important (even dominant) in many experimentally
inaccessible nuclei, so it is far from clear how the various calculations will
fare with data in the future.  In any event, the most striking fact is that all
the computations manage to existing data at roughly the same level of precision.
It may not be possible to do much better without moving beyond Skyrme QRPA, at
least while using a global parameter set as we have done here.

\subsection{Extrapolation to neutron-rich isotopes}

\begin{figure}
  \includegraphics[width=\columnwidth]{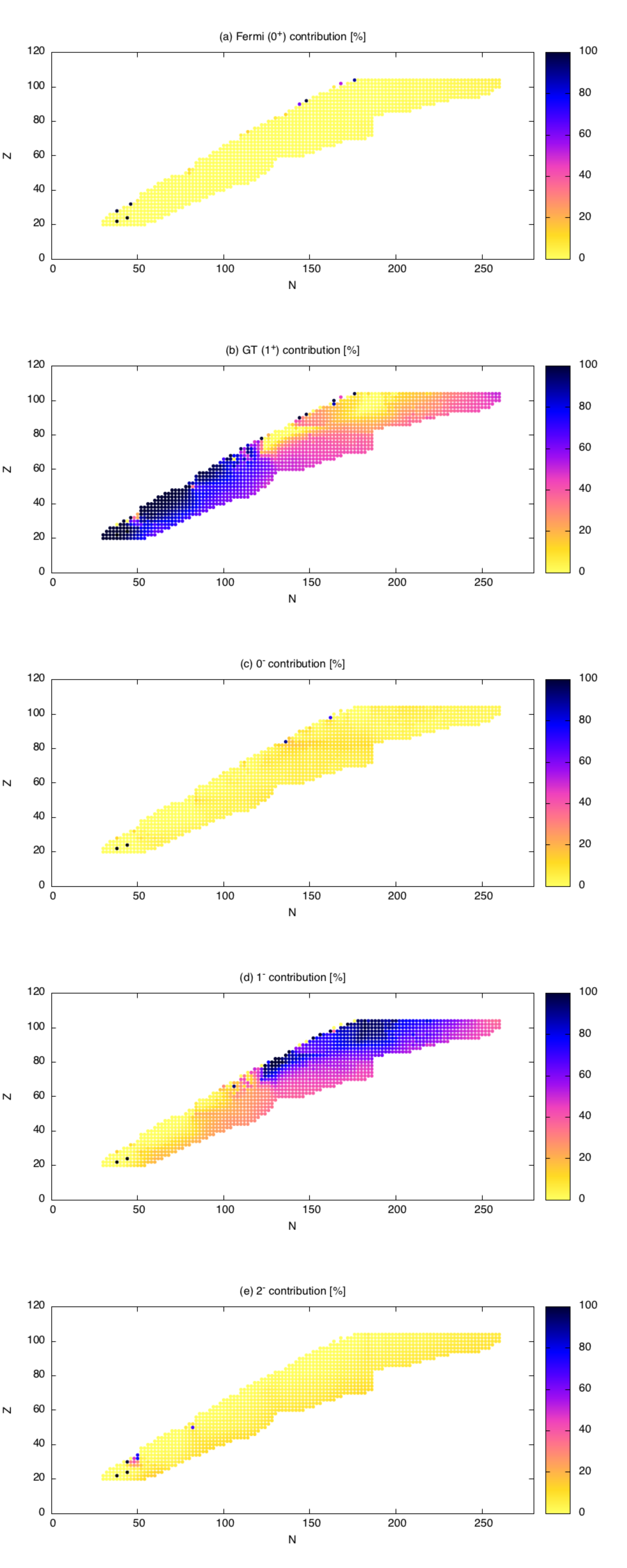}
  \caption{The contributions of different allowed and first-forbidden multipoles
  to the total computed beta-decay rates.  Only even-even nuclei are considered.
  }\label{fig:multipole-decomp}
\end{figure}

Figure~\ref{fig:multipole-decomp} displays the relative contribution to the
decay rate from each multipole.  Except in the immediate vicinity of the valley
of stability, the changes appear quite gradual as a function of $Z$ and $N$.  In
nuclei with large $Q$ values, the details of single-particle structure are less
important than in isotopes for which transitions to only a few low-energy states
are possible.

\begin{table*}[t]
  \begin{ruledtabular}
\begin{tabular}{lcccccccc}
$\mathcal{O}$ &    $d\mathcal{O}/dC^s_1$ & $d\mathcal{O}/dV_0$ & $d\mathcal{O}/dC^F_1$ & $d\mathcal{O}/dC^T_1$ & $d\mathcal{O}/dC^{\nabla s}_1$ & $d\mathcal{O}/dC^{\Delta s}_1$ & $d\mathcal{O}/dC^j_1$ & $d\mathcal{O}/dC^{\nabla j}_1$ \\
\hline
$^{208}$Pb $E_\textrm{GTR}$ & 57.261 & -0.000 &  2.434 & 5.869 &  0.429 & -1.002 &  0.000 &  0.143 \\
$^{112}$Sn $E_\textrm{GTR}$ & 29.498 & -1.032 &  1.432 & 2.863 &  0.286 & -0.573 &  0.000 &  0.000 \\
$^{76}$Ge $E_\textrm{GTR}$  & 45.115 & -7.225 &  2.004 & 4.295 &  0.429 & -1.145 &  0.000 &  0.000 \\
$^{130}$Te $E_\textrm{GTR}$ & 53.790 & -3.096 &  2.434 & 5.297 &  0.429 & -1.002 &  0.143 &  0.000 \\
$^{90}$Zr $E_\textrm{GTR}$  & 29.498 & -1.032 &  1.288 & 2.720 &  0.429 & -1.002 & -0.143 &  0.143 \\
$^{48}$Ca $E_\textrm{GTR}$  & 32.968 & -0.000 &  1.432 & 3.149 &  0.573 & -1.288 &  0.000 &  0.000 \\
\hline
$^{208}$Pb $E_\textrm{SDR}$ & 52.055 & -0.000 &  2.291 & 4.008 &  0.286 & -1.575 & -0.143 & -0.143 \\
$^{90}$Zr $E_\textrm{SDR}$  & 29.498 & -0.000 &  1.575 & 2.004 &  0.286 & -1.432 & -0.286 & -0.143 \\
\hline
$^{58}$Ti $\lg t$   &  4.749 & -4.318 &  0.203 & 0.445 &  0.045 & -0.109 & -0.011 & -0.002 \\
$^{78}$Zn $\lg t$   &  6.889 & -2.922 &  0.256 & 0.589 &  0.164 & -0.382 &  0.253 & -0.025 \\
$^{98}$Kr $\lg t$   &  5.410 & -3.252 &  0.265 & 0.559 &  0.050 & -0.116 & -0.012 & -0.003 \\
$^{126}$Cd $\lg t$  &  5.583 & -4.641 &  0.252 & 0.496 &  0.017 & -0.050 &  0.001 &  0.007 \\
$^{152}$Ce $\lg t$  &  5.409 & -2.474 &  0.293 & 0.540 &  0.051 & -0.120 &  0.003 & -0.009 \\
$^{166}$Gd $\lg t$  &  5.081 & -2.924 &  0.250 & 0.497 &  0.035 & -0.132 & -0.007 & -0.010 \\
$^{204}$Pt $\lg t$  &  3.755 & -3.340 & -0.015 & 0.160 & -0.018 & -0.316 & -0.076 &  0.026 \\
\end{tabular}
  \end{ruledtabular}
  \caption{The Jacobian matrix, evaluated at the result of the two-parameter fit
  1E.  All parameters except for the strength of isoscalar pairing are expressed
  in natural units. The strength of isoscalar pairing has been scaled by the
  strength of isovector pairing.  The derivatives of the $\lg t$ values are
  hence dimensionless and those of the resonance energies are in the units of
  MeV.}\label{tab:jacobian} 
  \end{table*}

\begin{table*}[bt]
  \begin{ruledtabular}
  \begin{tabular}{lcccl}
    fit & starting point & target set & $Q$ values & fitted parameters \\
    \hline
    1A & SkO'   & A & comp. & $V_0 = -173.176$, $C^s_1 = 128.279$ \\
    1B & SkO'   & B & comp. & $V_0 = -176.614$, $C^s_1 = 133.038$ \\
    1C & SkO'   & C & comp. & $V_0 = -176.097$, $C^s_1 = 126.966$ \\
    1D & SkO'   & E & comp. & $V_0 = -209.384$, $C^s_1 = 129.297$ \\
    1E & SkO'   & E & exp.  & $V_0 = -159.397$, $C^s_1 = 99.8479$ \\
    2  & SV-min & D & comp. & $V_0 = -165.567$, $C^s_1 = 132.271$ \\
    3A & SkO'   & E & comp. & $V_0 = -195.174$, $C^s_1 = 144.833$, $C^T_1 = -20.1618$, $C^F_1 = -10.3125$ \\
    3B & SkO'   & E & exp.  & $V_0 = -165.158$, $C^s_1 = 120.27$, $C^T_1 = -17.7435$, $C^F_1 = -17.9902$ \\
    4  & fit 3A & E & comp. & $C^j_1 = 54.5$, $C^{\nabla j}_1 = -78.7965$, $C^{\nabla s}_1 = -87.5$ \\
    5  & SkO'   & E & comp. & $V_0 = -191.875$, $C^s_1 = 146.182$, $C^j_1 = -86.4276$ \\
  \end{tabular}
  \end{ruledtabular}
  \caption{Summary of the various fits in this work. The functional listed for
  each fit dictates the values of the coupling constants that are not fit,
  except for that of $C^{\Delta s}_1$, which is set to zero everywhere to
  avoid finite-size instabilities. The units of $V_0$ and $C^s_1$ are
  MeV~fm$^3$, and the units of the other coupling constants are
  MeV~fm$^5$.}\label{tab:fits}
\end{table*}

Fig.~\ref{fig:multipole-decomp} also demonstrates the importance of going beyond
the allowed approximation.  In many heavy nuclei, the computed rates are
dominated by the first-forbidden channel.  Towards the drip line, both allowed
and forbidden channels are important for all masses.  The figure also shows that
the non-unique $1^-$ channel is usually the most important of the forbidden
multipoles.  Thus any quenching of the (unique) $2^-$ channel and anti-quenching
of the (non-unique) $0^-$ channel from meson-exchange currents
\cite{Warburton1991} would not have a significant impact on our overall results.
In the $1^-$ channel the contributions of several different operators makes the
effects of quenching hard to estimate. 

\begin{figure}[t]
\vspace{.3cm}
  \includegraphics[width=\columnwidth]{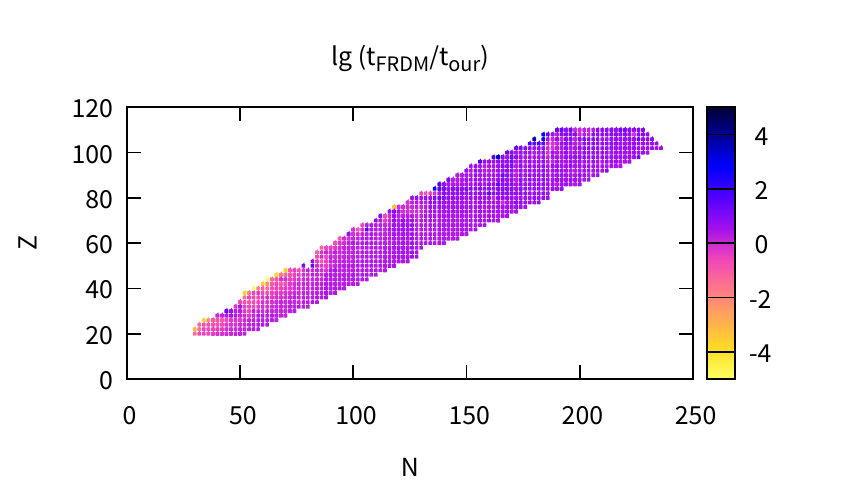}
  \caption{Comparison of our computed half-lives in neutron-rich nuclei
  with those of Ref.\ \cite{Moller2003} }\label{fig:our_vs_moller}.
\end{figure}

In Figure~\ref{fig:our_vs_moller} we compare our half-lives to those of M\"oller
\emph{et al.} \cite{Moller2003} in all medium and heavy even-even isotopes.  Our
half-lives tend to be longer then those of Ref.\ \cite{Moller2003} close to the
valley of stability in light nuclei and somewhat
shorter in heavy nuclei (with significant forbidden contributions).  Approaching
the neutron drip line, the two computations yield similar results up to a
constant offset in those of Ref.\ \cite{Moller2003} in even-even nuclei.  All
models can expect to do better near the drip line, where a significant fraction
of the total $\beta$-decay strength can be below threshold.

\begin{figure}[b]
  \includegraphics[width=\columnwidth]{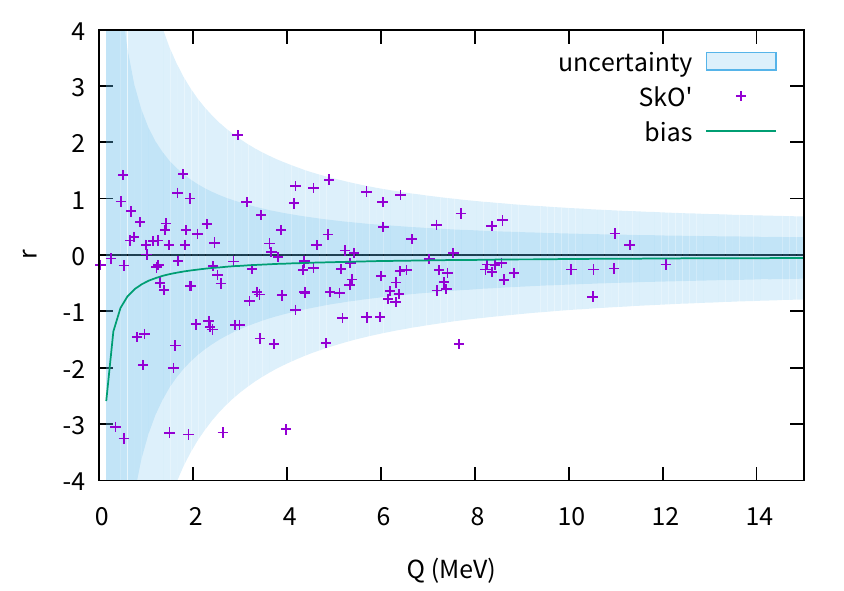}
  \caption{The fit 3A with the mean (green line) and one- and
  two-standard-deviation bands from our uncertainty model.
  }\label{fig:bands}
\end{figure}

Because our na\"ive model for uncertainties is based on several assumptions that
are only approximately correct or cannot easily be verified, we check its
predictions where there are enough data to do so.  Figure \ref{fig:bands} shows
the ratios of our half-lives to those of experiment together with the
uncertainty model's mean value and one- and two-standard-deviation bands, all as
a function of ground-state $Q$ value.  We can discount the model at very low $Q$
but it appears to work well above $Q \approx 4$ MeV.  Out of the 72 nuclei, 48
(66.7\%) fall within one standard deviation of the mean and 71 (98.6\%) within
two.  These numbers are consistent with what one would expect from a normal
distribution.  The model quantifies our statement above that calculations are
more accurate close to the drip line, where $Q$ is generally large.

\begin{figure}[t]
  \includegraphics[width=\columnwidth]{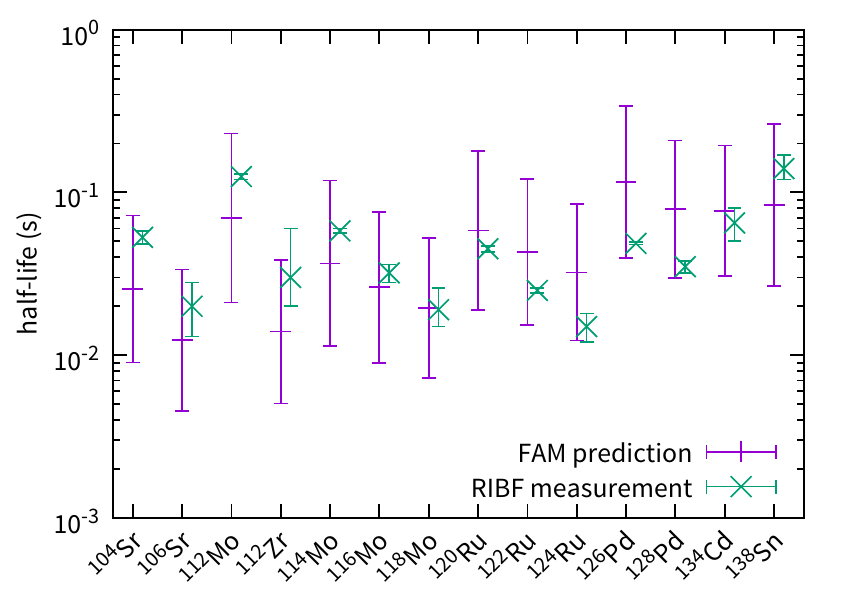}
  \caption{ Our predictions for the $14$ half-lives of neutron-rich even-even
  nuclei measured only recently \cite{Lorusso2015} and not included in the
  ENSDF data set.  All the measured half-lives fall within our one-sigma error
  bars, suggesting that our uncertainty estimates are too pessimistic
  in this particular region.
  }\label{fig:ribf}
\end{figure}

A recent RIBF measurement \cite{Lorusso2015} of 110 neutron-rich isotopes, 40 of
them previously unknown, allows us to test the reliability of our predictions
and especially our model for theoretical uncertainties.  Because the data are so
recent, we did not include them in any of our fits, and hence we are effectively
using older data to predict the results of these new measurements.  We have $28$
even-even nuclei with which to compare rates; for half of these there are
earlier data in the ENSDF set (Figure~\ref{fig:ribf}).  Our predictions agree
with experiment to within our theoretical uncertainty (though our error bars
may be a bit too large here).  Our uncertainty model thus appears to be
reasonable. 

\section{Conclusions}\label{sec:conclusions}

We have explored the ability of the axially-deformed Skyrme QRPA to provide a
global description of beta-decay rates in even-even neutron-rich nuclei.  With
experimental rates and charge-exchange resonance energies as fitting targets we
have found that among time-odd couplings, only those multiplying the
isoscalar-pairing and the spin-density parts of the functional are well
constrained; attempts to fit more than these two constants lead to overfitting.
The tensor contributions to the energy-density functional are, in particular,
not well constrained by this data.  To get more accurate Skyrme-QRPA
predictions, one can resort to local fits, i.e.\ $A$-dependent couplings.  The
recent work of Ref.\ \cite{Marketin2015}, for example, attaches a sensible $A$
dependence to the strength of isoscalar pairing.

The level of agreement between our calculations and data throughout the isotopic
chart is similar to that produced by other recent computations, in spite of our
consistent inclusion of deformation, tensor terms in the functional, etc.  It
could be difficult to do much better without an account of multiphonon effects,
as in the work, e.g., of Ref.\ \cite{Litvinova2014}.

The most glaring shortcoming of our work here is the restriction to even-even
nuclei.  An extension of the FAM to odd-mass nuclei will be the subject of a
future publication \cite{Shafer}. For the moment, we make our results for the
1387 even-even neutron-rich nuclei, with crudely estimated theoretical
uncertainties, available as supplementary material to this article.

\begin{acknowledgments}
We thank W.\ Nazarewicz, J.\ Dobaczewski, and S.\ Wild for useful discussions.
Support for this work was provided through the Scientific Discovery through
Advanced Computing (SciDAC) program funded by US Department of Energy, Office of
Science, Advanced Scientific Computing Research and Nuclear Physics, under
Contract No. DE-SC0008641, ER41896. This work used the Extreme Science and
Engineering Discovery Environment (XSEDE), which is supported by National
Science Foundation grant number ACI-1053575.
\end{acknowledgments}

\bibliographystyle{apsrev4-1}

\begin{thebibliography}{46}%
\makeatletter
\providecommand \@ifxundefined [1]{%
 \@ifx{#1\undefined}
}%
\providecommand \@ifnum [1]{%
 \ifnum #1\expandafter \@firstoftwo
 \else \expandafter \@secondoftwo
 \fi
}%
\providecommand \@ifx [1]{%
 \ifx #1\expandafter \@firstoftwo
 \else \expandafter \@secondoftwo
 \fi
}%
\providecommand \natexlab [1]{#1}%
\providecommand \enquote  [1]{``#1''}%
\providecommand \bibnamefont  [1]{#1}%
\providecommand \bibfnamefont [1]{#1}%
\providecommand \citenamefont [1]{#1}%
\providecommand \href@noop [0]{\@secondoftwo}%
\providecommand \href [0]{\begingroup \@sanitize@url \@href}%
\providecommand \@href[1]{\@@startlink{#1}\@@href}%
\providecommand \@@href[1]{\endgroup#1\@@endlink}%
\providecommand \@sanitize@url [0]{\catcode `\\12\catcode `\$12\catcode
  `\&12\catcode `\#12\catcode `\^12\catcode `\_12\catcode `\%12\relax}%
\providecommand \@@startlink[1]{}%
\providecommand \@@endlink[0]{}%
\providecommand \url  [0]{\begingroup\@sanitize@url \@url }%
\providecommand \@url [1]{\endgroup\@href {#1}{\urlprefix }}%
\providecommand \urlprefix  [0]{URL }%
\providecommand \Eprint [0]{\href }%
\providecommand \doibase [0]{http://dx.doi.org/}%
\providecommand \selectlanguage [0]{\@gobble}%
\providecommand \bibinfo  [0]{\@secondoftwo}%
\providecommand \bibfield  [0]{\@secondoftwo}%
\providecommand \translation [1]{[#1]}%
\providecommand \BibitemOpen [0]{}%
\providecommand \bibitemStop [0]{}%
\providecommand \bibitemNoStop [0]{.\EOS\space}%
\providecommand \EOS [0]{\spacefactor3000\relax}%
\providecommand \BibitemShut  [1]{\csname bibitem#1\endcsname}%
\let\auto@bib@innerbib\@empty
\bibitem [{\citenamefont {Mention}\ \emph {et~al.}(2011)\citenamefont
  {Mention}, \citenamefont {Fechner}, \citenamefont {Lasserre}, \citenamefont
  {Mueller}, \citenamefont {Lhuillier}, \citenamefont {Cribier},\ and\
  \citenamefont {Letourneau}}]{Mention2011}%
  \BibitemOpen
  \bibfield  {author} {\bibinfo {author} {\bibfnamefont {G.}~\bibnamefont
  {Mention}}, \bibinfo {author} {\bibfnamefont {M.}~\bibnamefont {Fechner}},
  \bibinfo {author} {\bibfnamefont {T.}~\bibnamefont {Lasserre}}, \bibinfo
  {author} {\bibfnamefont {T.~A.}\ \bibnamefont {Mueller}}, \bibinfo {author}
  {\bibfnamefont {D.}~\bibnamefont {Lhuillier}}, \bibinfo {author}
  {\bibfnamefont {M.}~\bibnamefont {Cribier}}, \ and\ \bibinfo {author}
  {\bibfnamefont {A.}~\bibnamefont {Letourneau}},\ }\href@noop {} {\bibfield
  {journal} {\bibinfo  {journal} {Phys. Rev. D}\ }\textbf {\bibinfo {volume}
  {83}},\ \bibinfo {pages} {073006} (\bibinfo {year} {2011})}\BibitemShut
  {NoStop}%
\bibitem [{\citenamefont {Hayes}\ \emph {et~al.}(2014)\citenamefont {Hayes},
  \citenamefont {Friar}, \citenamefont {Garvey}, \citenamefont {Jungman},\ and\
  \citenamefont {Jonkmans}}]{Hayes2014}%
  \BibitemOpen
  \bibfield  {author} {\bibinfo {author} {\bibfnamefont {A.~C.}\ \bibnamefont
  {Hayes}}, \bibinfo {author} {\bibfnamefont {J.~L.}\ \bibnamefont {Friar}},
  \bibinfo {author} {\bibfnamefont {G.~T.}\ \bibnamefont {Garvey}}, \bibinfo
  {author} {\bibfnamefont {G.}~\bibnamefont {Jungman}}, \ and\ \bibinfo
  {author} {\bibfnamefont {G.}~\bibnamefont {Jonkmans}},\ }\href@noop {}
  {\bibfield  {journal} {\bibinfo  {journal} {Phys. Rev. Lett.}\ }\textbf
  {\bibinfo {volume} {112}},\ \bibinfo {pages} {202501} (\bibinfo {year}
  {2014})}\BibitemShut {NoStop}%
\bibitem [{\citenamefont {Homma}\ \emph {et~al.}(1996)\citenamefont {Homma},
  \citenamefont {Bender}, \citenamefont {Hirsch}, \citenamefont {Muto},
  \citenamefont {Klapdor-Kleingrothaus},\ and\ \citenamefont
  {Oda}}]{Homma1996}%
  \BibitemOpen
  \bibfield  {author} {\bibinfo {author} {\bibfnamefont {H.}~\bibnamefont
  {Homma}}, \bibinfo {author} {\bibfnamefont {E.}~\bibnamefont {Bender}},
  \bibinfo {author} {\bibfnamefont {M.}~\bibnamefont {Hirsch}}, \bibinfo
  {author} {\bibfnamefont {K.}~\bibnamefont {Muto}}, \bibinfo {author}
  {\bibfnamefont {H.~V.}\ \bibnamefont {Klapdor-Kleingrothaus}}, \ and\
  \bibinfo {author} {\bibfnamefont {T.}~\bibnamefont {Oda}},\ }\href@noop {}
  {\bibfield  {journal} {\bibinfo  {journal} {Phys. Rev. C}\ }\textbf {\bibinfo
  {volume} {54}},\ \bibinfo {pages} {2972} (\bibinfo {year}
  {1996})}\BibitemShut {NoStop}%
\bibitem [{\citenamefont {M{\"o}ller}\ \emph {et~al.}(1997)\citenamefont
  {M{\"o}ller}, \citenamefont {Nix},\ and\ \citenamefont {Kratz}}]{Moller1997}%
  \BibitemOpen
  \bibfield  {author} {\bibinfo {author} {\bibfnamefont {P.}~\bibnamefont
  {M{\"o}ller}}, \bibinfo {author} {\bibfnamefont {J.~R.}\ \bibnamefont {Nix}},
  \ and\ \bibinfo {author} {\bibfnamefont {K.~L.}\ \bibnamefont {Kratz}},\
  }\href@noop {} {\bibfield  {journal} {\bibinfo  {journal} {Atomic Data and
  Nuclear Data Tables}\ }\textbf {\bibinfo {volume} {66}},\ \bibinfo {pages}
  {131} (\bibinfo {year} {1997})}\BibitemShut {NoStop}%
\bibitem [{\citenamefont {M{\"o}ller}\ \emph {et~al.}(2003)\citenamefont
  {M{\"o}ller}, \citenamefont {Pfeiffer},\ and\ \citenamefont
  {Kratz}}]{Moller2003}%
  \BibitemOpen
  \bibfield  {author} {\bibinfo {author} {\bibfnamefont {P.}~\bibnamefont
  {M{\"o}ller}}, \bibinfo {author} {\bibfnamefont {B.}~\bibnamefont
  {Pfeiffer}}, \ and\ \bibinfo {author} {\bibfnamefont {K.-L.}\ \bibnamefont
  {Kratz}},\ }\href@noop {} {\bibfield  {journal} {\bibinfo  {journal} {Phys.
  Rev. C}\ }\textbf {\bibinfo {volume} {67}},\ \bibinfo {pages} {055802}
  (\bibinfo {year} {2003})}\BibitemShut {NoStop}%
\bibitem [{\citenamefont {Nakata}\ \emph {et~al.}(1997)\citenamefont {Nakata},
  \citenamefont {Tachibana},\ and\ \citenamefont {Yamada}}]{Nakata1997}%
  \BibitemOpen
  \bibfield  {author} {\bibinfo {author} {\bibfnamefont {H.}~\bibnamefont
  {Nakata}}, \bibinfo {author} {\bibfnamefont {T.}~\bibnamefont {Tachibana}}, \
  and\ \bibinfo {author} {\bibfnamefont {M.}~\bibnamefont {Yamada}},\
  }\href@noop {} {\bibfield  {journal} {\bibinfo  {journal} {Nucl. Phys. A}\
  }\textbf {\bibinfo {volume} {625}},\ \bibinfo {pages} {521} (\bibinfo {year}
  {1997})}\BibitemShut {NoStop}%
\bibitem [{\citenamefont {Borzov}\ and\ \citenamefont
  {Goriely}(2000)}]{Borzov2000}%
  \BibitemOpen
  \bibfield  {author} {\bibinfo {author} {\bibfnamefont {I.~N.}\ \bibnamefont
  {Borzov}}\ and\ \bibinfo {author} {\bibfnamefont {S.}~\bibnamefont
  {Goriely}},\ }\href@noop {} {\bibfield  {journal} {\bibinfo  {journal} {Phys.
  Rev. C}\ }\textbf {\bibinfo {volume} {62}},\ \bibinfo {pages} {035501}
  (\bibinfo {year} {2000})}\BibitemShut {NoStop}%
\bibitem [{\citenamefont {Costiris}\ \emph {et~al.}(2009)\citenamefont
  {Costiris}, \citenamefont {Mavrommatis}, \citenamefont {Gernoth},\ and\
  \citenamefont {Clark}}]{Costiris2009}%
  \BibitemOpen
  \bibfield  {author} {\bibinfo {author} {\bibfnamefont {N.~J.}\ \bibnamefont
  {Costiris}}, \bibinfo {author} {\bibfnamefont {E.}~\bibnamefont
  {Mavrommatis}}, \bibinfo {author} {\bibfnamefont {K.~A.}\ \bibnamefont
  {Gernoth}}, \ and\ \bibinfo {author} {\bibfnamefont {J.~W.}\ \bibnamefont
  {Clark}},\ }\href@noop {} {\bibfield  {journal} {\bibinfo  {journal} {Phys.
  Rev. C}\ }\textbf {\bibinfo {volume} {80}},\ \bibinfo {pages} {044332}
  (\bibinfo {year} {2009})}\BibitemShut {NoStop}%
\bibitem [{\citenamefont {Marketin}\ \emph {et~al.}(2015)\citenamefont
  {Marketin}, \citenamefont {Huther},\ and\ \citenamefont
  {Mart{\'\i}nez-Pinedo}}]{Marketin2015}%
  \BibitemOpen
  \bibfield  {author} {\bibinfo {author} {\bibfnamefont {T.}~\bibnamefont
  {Marketin}}, \bibinfo {author} {\bibfnamefont {L.}~\bibnamefont {Huther}}, \
  and\ \bibinfo {author} {\bibfnamefont {G.}~\bibnamefont
  {Mart{\'\i}nez-Pinedo}},\ }\href@noop {} {} (\bibinfo {year} {2015}),\
  \bibinfo {note} {arXiv:1507.07442}\BibitemShut {NoStop}%
\bibitem [{\citenamefont {Fang}\ \emph {et~al.}(2013)\citenamefont {Fang},
  \citenamefont {Brown},\ and\ \citenamefont {Suzuki}}]{Fang2013}%
  \BibitemOpen
  \bibfield  {author} {\bibinfo {author} {\bibfnamefont {D.-L.}\ \bibnamefont
  {Fang}}, \bibinfo {author} {\bibfnamefont {B.~A.}\ \bibnamefont {Brown}}, \
  and\ \bibinfo {author} {\bibfnamefont {T.}~\bibnamefont {Suzuki}},\
  }\href@noop {} {\bibfield  {journal} {\bibinfo  {journal} {Phys. Rev. C}\
  }\textbf {\bibinfo {volume} {88}},\ \bibinfo {pages} {024314} (\bibinfo
  {year} {2013})}\BibitemShut {NoStop}%
\bibitem [{\citenamefont {Ni}\ and\ \citenamefont
  {Ren}(2014{\natexlab{a}})}]{Ni2014}%
  \BibitemOpen
  \bibfield  {author} {\bibinfo {author} {\bibfnamefont {D.}~\bibnamefont
  {Ni}}\ and\ \bibinfo {author} {\bibfnamefont {Z.}~\bibnamefont {Ren}},\
  }\href@noop {} {\bibfield  {journal} {\bibinfo  {journal} {Phys. Rev. C}\
  }\textbf {\bibinfo {volume} {89}},\ \bibinfo {pages} {064320} (\bibinfo
  {year} {2014}{\natexlab{a}})}\BibitemShut {NoStop}%
\bibitem [{\citenamefont {Ni}\ and\ \citenamefont
  {Ren}(2014{\natexlab{b}})}]{Ni2014a}%
  \BibitemOpen
  \bibfield  {author} {\bibinfo {author} {\bibfnamefont {D.}~\bibnamefont
  {Ni}}\ and\ \bibinfo {author} {\bibfnamefont {Z.}~\bibnamefont {Ren}},\
  }\href@noop {} {\bibfield  {journal} {\bibinfo  {journal} {J. Phys. G: Nucl.
  Part. Phys.}\ }\textbf {\bibinfo {volume} {41}},\ \bibinfo {pages} {125102}
  (\bibinfo {year} {2014}{\natexlab{b}})}\BibitemShut {NoStop}%
\bibitem [{\citenamefont {Martini}\ \emph {et~al.}(2014)\citenamefont
  {Martini}, \citenamefont {P{\'e}ru},\ and\ \citenamefont
  {Goriely}}]{Martini2014}%
  \BibitemOpen
  \bibfield  {author} {\bibinfo {author} {\bibfnamefont {M.}~\bibnamefont
  {Martini}}, \bibinfo {author} {\bibfnamefont {S.}~\bibnamefont {P{\'e}ru}}, \
  and\ \bibinfo {author} {\bibfnamefont {S.}~\bibnamefont {Goriely}},\
  }\href@noop {} {\bibfield  {journal} {\bibinfo  {journal} {Phys. Rev. C}\
  }\textbf {\bibinfo {volume} {89}},\ \bibinfo {pages} {044306} (\bibinfo
  {year} {2014})}\BibitemShut {NoStop}%
\bibitem [{\citenamefont {Niu}\ \emph {et~al.}(2013)\citenamefont {Niu},
  \citenamefont {Niu}, \citenamefont {Liang}, \citenamefont {Long},
  \citenamefont {Nik{\v s}i{\'c}}, \citenamefont {Vretenar},\ and\
  \citenamefont {Meng}}]{Niu2013}%
  \BibitemOpen
  \bibfield  {author} {\bibinfo {author} {\bibfnamefont {Z.~M.}\ \bibnamefont
  {Niu}}, \bibinfo {author} {\bibfnamefont {Y.~F.}\ \bibnamefont {Niu}},
  \bibinfo {author} {\bibfnamefont {H.~Z.}\ \bibnamefont {Liang}}, \bibinfo
  {author} {\bibfnamefont {W.~H.}\ \bibnamefont {Long}}, \bibinfo {author}
  {\bibfnamefont {T.}~\bibnamefont {Nik{\v s}i{\'c}}}, \bibinfo {author}
  {\bibfnamefont {D.}~\bibnamefont {Vretenar}}, \ and\ \bibinfo {author}
  {\bibfnamefont {J.}~\bibnamefont {Meng}},\ }\href@noop {} {\bibfield
  {journal} {\bibinfo  {journal} {Phys. Lett. B}\ }\textbf {\bibinfo {volume}
  {723}},\ \bibinfo {pages} {172} (\bibinfo {year} {2013})}\BibitemShut
  {NoStop}%
\bibitem [{\citenamefont {Nik{\v s}i{\'c}}\ \emph {et~al.}(2005)\citenamefont
  {Nik{\v s}i{\'c}}, \citenamefont {Marketin}, \citenamefont {Vretenar},
  \citenamefont {Paar},\ and\ \citenamefont {Ring}}]{Niksic2005}%
  \BibitemOpen
  \bibfield  {author} {\bibinfo {author} {\bibfnamefont {T.}~\bibnamefont
  {Nik{\v s}i{\'c}}}, \bibinfo {author} {\bibfnamefont {T.}~\bibnamefont
  {Marketin}}, \bibinfo {author} {\bibfnamefont {D.}~\bibnamefont {Vretenar}},
  \bibinfo {author} {\bibfnamefont {N.}~\bibnamefont {Paar}}, \ and\ \bibinfo
  {author} {\bibfnamefont {P.}~\bibnamefont {Ring}},\ }\href@noop {} {\bibfield
   {journal} {\bibinfo  {journal} {Phys. Rev. C}\ }\textbf {\bibinfo {volume}
  {71}},\ \bibinfo {pages} {014308} (\bibinfo {year} {2005})}\BibitemShut
  {NoStop}%
\bibitem [{\citenamefont {Mustonen}\ \emph {et~al.}(2014)\citenamefont
  {Mustonen}, \citenamefont {Shafer}, \citenamefont {Zenginerler},\ and\
  \citenamefont {Engel}}]{Mustonen2014}%
  \BibitemOpen
  \bibfield  {author} {\bibinfo {author} {\bibfnamefont {M.~T.}\ \bibnamefont
  {Mustonen}}, \bibinfo {author} {\bibfnamefont {T.}~\bibnamefont {Shafer}},
  \bibinfo {author} {\bibfnamefont {Z.}~\bibnamefont {Zenginerler}}, \ and\
  \bibinfo {author} {\bibfnamefont {J.}~\bibnamefont {Engel}},\ }\href@noop {}
  {\bibfield  {journal} {\bibinfo  {journal} {Phys. Rev. C}\ }\textbf {\bibinfo
  {volume} {90}},\ \bibinfo {pages} {024308} (\bibinfo {year}
  {2014})}\BibitemShut {NoStop}%
\bibitem [{\citenamefont {Minato}\ and\ \citenamefont
  {Bai}(2013)}]{Minato2013a}%
  \BibitemOpen
  \bibfield  {author} {\bibinfo {author} {\bibfnamefont {F.}~\bibnamefont
  {Minato}}\ and\ \bibinfo {author} {\bibfnamefont {C.~L.}\ \bibnamefont
  {Bai}},\ }\href@noop {} {\bibfield  {journal} {\bibinfo  {journal} {Phys.
  Rev. Lett.}\ }\textbf {\bibinfo {volume} {110}},\ \bibinfo {pages} {122501}
  (\bibinfo {year} {2013})}\BibitemShut {NoStop}%
\bibitem [{\citenamefont {De~Donno}\ \emph {et~al.}(2014)\citenamefont
  {De~Donno}, \citenamefont {Co'}, \citenamefont {Anguiano},\ and\
  \citenamefont {Lallena}}]{De-Donno2014}%
  \BibitemOpen
  \bibfield  {author} {\bibinfo {author} {\bibfnamefont {V.}~\bibnamefont
  {De~Donno}}, \bibinfo {author} {\bibfnamefont {G.}~\bibnamefont {Co'}},
  \bibinfo {author} {\bibfnamefont {M.}~\bibnamefont {Anguiano}}, \ and\
  \bibinfo {author} {\bibfnamefont {A.~M.}\ \bibnamefont {Lallena}},\
  }\href@noop {} {\bibfield  {journal} {\bibinfo  {journal} {Phys. Rev. C}\
  }\textbf {\bibinfo {volume} {90}},\ \bibinfo {pages} {024326} (\bibinfo
  {year} {2014})}\BibitemShut {NoStop}%
\bibitem [{\citenamefont {Nakatsukasa}\ \emph {et~al.}(2007)\citenamefont
  {Nakatsukasa}, \citenamefont {Inakura},\ and\ \citenamefont
  {Yabana}}]{Nakatsukasa2007}%
  \BibitemOpen
  \bibfield  {author} {\bibinfo {author} {\bibfnamefont {T.}~\bibnamefont
  {Nakatsukasa}}, \bibinfo {author} {\bibfnamefont {T.}~\bibnamefont
  {Inakura}}, \ and\ \bibinfo {author} {\bibfnamefont {K.}~\bibnamefont
  {Yabana}},\ }\href@noop {} {\bibfield  {journal} {\bibinfo  {journal} {Phys.
  Rev. C}\ }\textbf {\bibinfo {volume} {76}},\ \bibinfo {pages} {024318}
  (\bibinfo {year} {2007})}\BibitemShut {NoStop}%
\bibitem [{\citenamefont {Avogadro}\ and\ \citenamefont
  {Nakatsukasa}(2011)}]{Avogadro2011}%
  \BibitemOpen
  \bibfield  {author} {\bibinfo {author} {\bibfnamefont {P.}~\bibnamefont
  {Avogadro}}\ and\ \bibinfo {author} {\bibfnamefont {T.}~\bibnamefont
  {Nakatsukasa}},\ }\href@noop {} {\bibfield  {journal} {\bibinfo  {journal}
  {Phys. Rev. C}\ }\textbf {\bibinfo {volume} {84}},\ \bibinfo {pages} {014314}
  (\bibinfo {year} {2011})}\BibitemShut {NoStop}%
\bibitem [{\citenamefont {Nik{\v s}i{\'c}}\ \emph {et~al.}(2013)\citenamefont
  {Nik{\v s}i{\'c}}, \citenamefont {Kralj}, \citenamefont {Tuti{\v s}},
  \citenamefont {Vretenar},\ and\ \citenamefont {Ring}}]{Niksic2013}%
  \BibitemOpen
  \bibfield  {author} {\bibinfo {author} {\bibfnamefont {T.}~\bibnamefont
  {Nik{\v s}i{\'c}}}, \bibinfo {author} {\bibfnamefont {N.}~\bibnamefont
  {Kralj}}, \bibinfo {author} {\bibfnamefont {T.}~\bibnamefont {Tuti{\v s}}},
  \bibinfo {author} {\bibfnamefont {D.}~\bibnamefont {Vretenar}}, \ and\
  \bibinfo {author} {\bibfnamefont {P.}~\bibnamefont {Ring}},\ }\href@noop {}
  {\bibfield  {journal} {\bibinfo  {journal} {Phys. Rev. C}\ }\textbf {\bibinfo
  {volume} {88}},\ \bibinfo {pages} {044327} (\bibinfo {year}
  {2013})}\BibitemShut {NoStop}%
\bibitem [{\citenamefont {Hinohara}\ \emph {et~al.}(2015)\citenamefont
  {Hinohara}, \citenamefont {Kortelainen}, \citenamefont {Nazarewicz},\ and\
  \citenamefont {Olsen}}]{Hinohara2015}%
  \BibitemOpen
  \bibfield  {author} {\bibinfo {author} {\bibfnamefont {N.}~\bibnamefont
  {Hinohara}}, \bibinfo {author} {\bibfnamefont {M.}~\bibnamefont
  {Kortelainen}}, \bibinfo {author} {\bibfnamefont {W.}~\bibnamefont
  {Nazarewicz}}, \ and\ \bibinfo {author} {\bibfnamefont {E.}~\bibnamefont
  {Olsen}},\ }\href@noop {} {\bibfield  {journal} {\bibinfo  {journal} {Phys.
  Rev. C}\ }\textbf {\bibinfo {volume} {91}},\ \bibinfo {pages} {044323}
  (\bibinfo {year} {2015})}\BibitemShut {NoStop}%
\bibitem [{\citenamefont {Perli{\'n}ska}\ \emph {et~al.}(2004)\citenamefont
  {Perli{\'n}ska}, \citenamefont {Rohozi{\'n}ski}, \citenamefont
  {Dobaczewski},\ and\ \citenamefont {Nazarewicz}}]{Perlinska2004}%
  \BibitemOpen
  \bibfield  {author} {\bibinfo {author} {\bibfnamefont {E.}~\bibnamefont
  {Perli{\'n}ska}}, \bibinfo {author} {\bibfnamefont {S.~G.}\ \bibnamefont
  {Rohozi{\'n}ski}}, \bibinfo {author} {\bibfnamefont {J.}~\bibnamefont
  {Dobaczewski}}, \ and\ \bibinfo {author} {\bibfnamefont {W.}~\bibnamefont
  {Nazarewicz}},\ }\href@noop {} {\bibfield  {journal} {\bibinfo  {journal}
  {Phys. Rev. C}\ }\textbf {\bibinfo {volume} {69}},\ \bibinfo {pages} {014316}
  (\bibinfo {year} {2004})}\BibitemShut {NoStop}%
\bibitem [{\citenamefont {Bender}\ \emph {et~al.}(2002)\citenamefont {Bender},
  \citenamefont {Dobaczewski}, \citenamefont {Engel},\ and\ \citenamefont
  {Nazarewicz}}]{Bender2002}%
  \BibitemOpen
  \bibfield  {author} {\bibinfo {author} {\bibfnamefont {M.}~\bibnamefont
  {Bender}}, \bibinfo {author} {\bibfnamefont {J.}~\bibnamefont {Dobaczewski}},
  \bibinfo {author} {\bibfnamefont {J.}~\bibnamefont {Engel}}, \ and\ \bibinfo
  {author} {\bibfnamefont {W.}~\bibnamefont {Nazarewicz}},\ }\href@noop {}
  {\bibfield  {journal} {\bibinfo  {journal} {Phys. Rev. C}\ }\textbf {\bibinfo
  {volume} {65}},\ \bibinfo {pages} {054322} (\bibinfo {year}
  {2002})}\BibitemShut {NoStop}%
\bibitem [{\citenamefont {B{\"a}ckman}\ \emph {et~al.}(1979)\citenamefont
  {B{\"a}ckman}, \citenamefont {Sj{\"o}berg},\ and\ \citenamefont
  {Jackson}}]{Backman1979a}%
  \BibitemOpen
  \bibfield  {author} {\bibinfo {author} {\bibfnamefont {S.~O.}\ \bibnamefont
  {B{\"a}ckman}}, \bibinfo {author} {\bibfnamefont {O.}~\bibnamefont
  {Sj{\"o}berg}}, \ and\ \bibinfo {author} {\bibfnamefont {A.~D.}\ \bibnamefont
  {Jackson}},\ }\href@noop {} {\bibfield  {journal} {\bibinfo  {journal} {Nucl.
  Phys. A}\ }\textbf {\bibinfo {volume} {321}},\ \bibinfo {pages} {10}
  (\bibinfo {year} {1979})}\BibitemShut {NoStop}%
\bibitem [{\citenamefont {Dobaczewski}\ \emph {et~al.}(2014)\citenamefont
  {Dobaczewski}, \citenamefont {Nazarewicz},\ and\ \citenamefont
  {Reinhard}}]{Dobaczewski2014}%
  \BibitemOpen
  \bibfield  {author} {\bibinfo {author} {\bibfnamefont {J.}~\bibnamefont
  {Dobaczewski}}, \bibinfo {author} {\bibfnamefont {W.}~\bibnamefont
  {Nazarewicz}}, \ and\ \bibinfo {author} {\bibfnamefont {P.~G.}\ \bibnamefont
  {Reinhard}},\ }\href@noop {} {\bibfield  {journal} {\bibinfo  {journal} {J.
  Phys. G: Nucl. Part. Phys.}\ }\textbf {\bibinfo {volume} {41}},\ \bibinfo
  {pages} {074001} (\bibinfo {year} {2014})}\BibitemShut {NoStop}%
\bibitem [{\citenamefont {Primakoff}\ and\ \citenamefont
  {Rosen}(1959)}]{Primakoff1959}%
  \BibitemOpen
  \bibfield  {author} {\bibinfo {author} {\bibfnamefont {H.}~\bibnamefont
  {Primakoff}}\ and\ \bibinfo {author} {\bibfnamefont {S.~P.}\ \bibnamefont
  {Rosen}},\ }\href@noop {} {\bibfield  {journal} {\bibinfo  {journal} {Rep.
  Prog. Phys.}\ }\textbf {\bibinfo {volume} {22}},\ \bibinfo {pages} {121}
  (\bibinfo {year} {1959})}\BibitemShut {NoStop}%
\bibitem [{\citenamefont {Akimune}\ \emph {et~al.}(1995)\citenamefont
  {Akimune}, \citenamefont {Daito}, \citenamefont {Fujita}, \citenamefont
  {Fujiwara}, \citenamefont {Greenfield}, \citenamefont {Harakeh},
  \citenamefont {Inomata}, \citenamefont {J\"anecke}, \citenamefont {Katori},
  \citenamefont {Nakayama}, \citenamefont {Sakai}, \citenamefont {Sakemi},
  \citenamefont {Tanaka},\ and\ \citenamefont {Yosoi}}]{Akimune1995}%
  \BibitemOpen
  \bibfield  {author} {\bibinfo {author} {\bibfnamefont {H.}~\bibnamefont
  {Akimune}}, \bibinfo {author} {\bibfnamefont {I.}~\bibnamefont {Daito}},
  \bibinfo {author} {\bibfnamefont {Y.}~\bibnamefont {Fujita}}, \bibinfo
  {author} {\bibfnamefont {M.}~\bibnamefont {Fujiwara}}, \bibinfo {author}
  {\bibfnamefont {M.~B.}\ \bibnamefont {Greenfield}}, \bibinfo {author}
  {\bibfnamefont {M.~N.}\ \bibnamefont {Harakeh}}, \bibinfo {author}
  {\bibfnamefont {T.}~\bibnamefont {Inomata}}, \bibinfo {author} {\bibfnamefont
  {J.}~\bibnamefont {J\"anecke}}, \bibinfo {author} {\bibfnamefont
  {K.}~\bibnamefont {Katori}}, \bibinfo {author} {\bibfnamefont
  {S.}~\bibnamefont {Nakayama}}, \bibinfo {author} {\bibfnamefont
  {H.}~\bibnamefont {Sakai}}, \bibinfo {author} {\bibfnamefont
  {Y.}~\bibnamefont {Sakemi}}, \bibinfo {author} {\bibfnamefont
  {M.}~\bibnamefont {Tanaka}}, \ and\ \bibinfo {author} {\bibfnamefont
  {M.}~\bibnamefont {Yosoi}},\ }\href@noop {} {\bibfield  {journal} {\bibinfo
  {journal} {Phys. Rev. C}\ }\textbf {\bibinfo {volume} {52}},\ \bibinfo
  {pages} {604} (\bibinfo {year} {1995})}\BibitemShut {NoStop}%
\bibitem [{\citenamefont {Anderson}\ \emph {et~al.}(1985)\citenamefont
  {Anderson}, \citenamefont {Chittrakarn}, \citenamefont {Baldwin},
  \citenamefont {Lebo}, \citenamefont {Madey}, \citenamefont {Tandy},
  \citenamefont {Watson}, \citenamefont {Brown},\ and\ \citenamefont
  {Foster}}]{Anderson1985}%
  \BibitemOpen
  \bibfield  {author} {\bibinfo {author} {\bibfnamefont {B.~D.}\ \bibnamefont
  {Anderson}}, \bibinfo {author} {\bibfnamefont {T.}~\bibnamefont
  {Chittrakarn}}, \bibinfo {author} {\bibfnamefont {A.~R.}\ \bibnamefont
  {Baldwin}}, \bibinfo {author} {\bibfnamefont {C.}~\bibnamefont {Lebo}},
  \bibinfo {author} {\bibfnamefont {R.}~\bibnamefont {Madey}}, \bibinfo
  {author} {\bibfnamefont {P.~C.}\ \bibnamefont {Tandy}}, \bibinfo {author}
  {\bibfnamefont {J.~W.}\ \bibnamefont {Watson}}, \bibinfo {author}
  {\bibfnamefont {B.~A.}\ \bibnamefont {Brown}}, \ and\ \bibinfo {author}
  {\bibfnamefont {C.~C.}\ \bibnamefont {Foster}},\ }\href@noop {} {\bibfield
  {journal} {\bibinfo  {journal} {Phys. Rev. C}\ }\textbf {\bibinfo {volume}
  {31}},\ \bibinfo {pages} {1161} (\bibinfo {year} {1985})}\BibitemShut
  {NoStop}%
\bibitem [{\citenamefont {Madey}\ \emph {et~al.}(1989)\citenamefont {Madey},
  \citenamefont {Flanders}, \citenamefont {Anderson}, \citenamefont {Baldwin},
  \citenamefont {Watson}, \citenamefont {Austin}, \citenamefont {Foster},
  \citenamefont {Klapdor},\ and\ \citenamefont {Grotz}}]{Madey1989}%
  \BibitemOpen
  \bibfield  {author} {\bibinfo {author} {\bibfnamefont {R.}~\bibnamefont
  {Madey}}, \bibinfo {author} {\bibfnamefont {B.~S.}\ \bibnamefont {Flanders}},
  \bibinfo {author} {\bibfnamefont {B.~D.}\ \bibnamefont {Anderson}}, \bibinfo
  {author} {\bibfnamefont {A.~R.}\ \bibnamefont {Baldwin}}, \bibinfo {author}
  {\bibfnamefont {J.~W.}\ \bibnamefont {Watson}}, \bibinfo {author}
  {\bibfnamefont {S.~M.}\ \bibnamefont {Austin}}, \bibinfo {author}
  {\bibfnamefont {C.~C.}\ \bibnamefont {Foster}}, \bibinfo {author}
  {\bibfnamefont {H.~V.}\ \bibnamefont {Klapdor}}, \ and\ \bibinfo {author}
  {\bibfnamefont {K.}~\bibnamefont {Grotz}},\ }\href@noop {} {\bibfield
  {journal} {\bibinfo  {journal} {Phys. Rev. C}\ }\textbf {\bibinfo {volume}
  {40}},\ \bibinfo {pages} {540} (\bibinfo {year} {1989})}\BibitemShut
  {NoStop}%
\bibitem [{\citenamefont {Pham}\ \emph {et~al.}(1995)\citenamefont {Pham},
  \citenamefont {J\"anecke}, \citenamefont {Roberts}, \citenamefont {Harakeh},
  \citenamefont {Berg}, \citenamefont {Chang}, \citenamefont {Liu},
  \citenamefont {Stephenson}, \citenamefont {Davis}, \citenamefont {Akimune},\
  and\ \citenamefont {Fujiwara}}]{Pham1995}%
  \BibitemOpen
  \bibfield  {author} {\bibinfo {author} {\bibfnamefont {K.}~\bibnamefont
  {Pham}}, \bibinfo {author} {\bibfnamefont {J.}~\bibnamefont {J\"anecke}},
  \bibinfo {author} {\bibfnamefont {D.~A.}\ \bibnamefont {Roberts}}, \bibinfo
  {author} {\bibfnamefont {M.~N.}\ \bibnamefont {Harakeh}}, \bibinfo {author}
  {\bibfnamefont {G.~P.~A.}\ \bibnamefont {Berg}}, \bibinfo {author}
  {\bibfnamefont {S.}~\bibnamefont {Chang}}, \bibinfo {author} {\bibfnamefont
  {J.}~\bibnamefont {Liu}}, \bibinfo {author} {\bibfnamefont {E.~J.}\
  \bibnamefont {Stephenson}}, \bibinfo {author} {\bibfnamefont {B.~F.}\
  \bibnamefont {Davis}}, \bibinfo {author} {\bibfnamefont {H.}~\bibnamefont
  {Akimune}}, \ and\ \bibinfo {author} {\bibfnamefont {M.}~\bibnamefont
  {Fujiwara}},\ }\href@noop {} {\bibfield  {journal} {\bibinfo  {journal}
  {Phys. Rev. C}\ }\textbf {\bibinfo {volume} {51}},\ \bibinfo {pages} {526}
  (\bibinfo {year} {1995})}\BibitemShut {NoStop}%
\bibitem [{\citenamefont {Wakasa}\ \emph {et~al.}(1997)\citenamefont {Wakasa},
  \citenamefont {Sakai}, \citenamefont {Okamura}, \citenamefont {Otsu},
  \citenamefont {Fujita}, \citenamefont {Ishida}, \citenamefont {Sakamoto},
  \citenamefont {Uesaka}, \citenamefont {Satou}, \citenamefont {Greenfield},\
  and\ \citenamefont {Hatanaka}}]{Wakasa1997}%
  \BibitemOpen
  \bibfield  {author} {\bibinfo {author} {\bibfnamefont {T.}~\bibnamefont
  {Wakasa}}, \bibinfo {author} {\bibfnamefont {H.}~\bibnamefont {Sakai}},
  \bibinfo {author} {\bibfnamefont {H.}~\bibnamefont {Okamura}}, \bibinfo
  {author} {\bibfnamefont {H.}~\bibnamefont {Otsu}}, \bibinfo {author}
  {\bibfnamefont {S.}~\bibnamefont {Fujita}}, \bibinfo {author} {\bibfnamefont
  {S.}~\bibnamefont {Ishida}}, \bibinfo {author} {\bibfnamefont
  {N.}~\bibnamefont {Sakamoto}}, \bibinfo {author} {\bibfnamefont
  {T.}~\bibnamefont {Uesaka}}, \bibinfo {author} {\bibfnamefont
  {Y.}~\bibnamefont {Satou}}, \bibinfo {author} {\bibfnamefont {M.~B.}\
  \bibnamefont {Greenfield}}, \ and\ \bibinfo {author} {\bibfnamefont
  {K.}~\bibnamefont {Hatanaka}},\ }\href@noop {} {\bibfield  {journal}
  {\bibinfo  {journal} {Phys. Rev. C}\ }\textbf {\bibinfo {volume} {55}},\
  \bibinfo {pages} {2909} (\bibinfo {year} {1997})}\BibitemShut {NoStop}%
\bibitem [{\citenamefont {Yako}\ \emph {et~al.}(2006)\citenamefont {Yako},
  \citenamefont {Sagawa},\ and\ \citenamefont {Sakai}}]{Yako2006}%
  \BibitemOpen
  \bibfield  {author} {\bibinfo {author} {\bibfnamefont {K.}~\bibnamefont
  {Yako}}, \bibinfo {author} {\bibfnamefont {H.}~\bibnamefont {Sagawa}}, \ and\
  \bibinfo {author} {\bibfnamefont {H.}~\bibnamefont {Sakai}},\ }\href@noop {}
  {\bibfield  {journal} {\bibinfo  {journal} {Phys. Rev. C}\ }\textbf {\bibinfo
  {volume} {74}},\ \bibinfo {pages} {051303(R)} (\bibinfo {year}
  {2006})}\BibitemShut {NoStop}%
\bibitem [{\citenamefont {Wakasa}\ \emph {et~al.}(2012)\citenamefont {Wakasa},
  \citenamefont {Okamoto}, \citenamefont {Dozono}, \citenamefont {Hatanaka},
  \citenamefont {Ichimura}, \citenamefont {Kuroita}, \citenamefont {Maeda},
  \citenamefont {Miyasako}, \citenamefont {Noro}, \citenamefont {Saito},
  \citenamefont {Sakemi}, \citenamefont {Yabe},\ and\ \citenamefont
  {Yako}}]{Wakasa2012}%
  \BibitemOpen
  \bibfield  {author} {\bibinfo {author} {\bibfnamefont {T.}~\bibnamefont
  {Wakasa}}, \bibinfo {author} {\bibfnamefont {M.}~\bibnamefont {Okamoto}},
  \bibinfo {author} {\bibfnamefont {M.}~\bibnamefont {Dozono}}, \bibinfo
  {author} {\bibfnamefont {K.}~\bibnamefont {Hatanaka}}, \bibinfo {author}
  {\bibfnamefont {M.}~\bibnamefont {Ichimura}}, \bibinfo {author}
  {\bibfnamefont {S.}~\bibnamefont {Kuroita}}, \bibinfo {author} {\bibfnamefont
  {Y.}~\bibnamefont {Maeda}}, \bibinfo {author} {\bibfnamefont
  {H.}~\bibnamefont {Miyasako}}, \bibinfo {author} {\bibfnamefont
  {T.}~\bibnamefont {Noro}}, \bibinfo {author} {\bibfnamefont {T.}~\bibnamefont
  {Saito}}, \bibinfo {author} {\bibfnamefont {Y.}~\bibnamefont {Sakemi}},
  \bibinfo {author} {\bibfnamefont {T.}~\bibnamefont {Yabe}}, \ and\ \bibinfo
  {author} {\bibfnamefont {K.}~\bibnamefont {Yako}},\ }\href@noop {} {\bibfield
   {journal} {\bibinfo  {journal} {Phys. Rev. C}\ }\textbf {\bibinfo {volume}
  {85}},\ \bibinfo {pages} {064606} (\bibinfo {year} {2012})}\BibitemShut
  {NoStop}%
\bibitem [{ens(2014)}]{ensdf2014apr}%
  \BibitemOpen
  \href {http://nndc.bnl.gov/ensdf} {\enquote {\bibinfo {title} {Evaluated
  nuclear structure data file ({ENSDF})},}\ }\bibinfo {howpublished}
  {http://nndc.bnl.gov/ensdf} (\bibinfo {year} {2014})\BibitemShut {NoStop}%
\bibitem [{\citenamefont {Stoitsov}\ \emph {et~al.}(2013)\citenamefont
  {Stoitsov}, \citenamefont {Schunck}, \citenamefont {Kortelainen},
  \citenamefont {Michel}, \citenamefont {Nam}, \citenamefont {Olsen},
  \citenamefont {Sarich},\ and\ \citenamefont {Wild}}]{Stoitsov2013}%
  \BibitemOpen
  \bibfield  {author} {\bibinfo {author} {\bibfnamefont {M.~V.}\ \bibnamefont
  {Stoitsov}}, \bibinfo {author} {\bibfnamefont {N.}~\bibnamefont {Schunck}},
  \bibinfo {author} {\bibfnamefont {M.}~\bibnamefont {Kortelainen}}, \bibinfo
  {author} {\bibfnamefont {N.}~\bibnamefont {Michel}}, \bibinfo {author}
  {\bibfnamefont {H.}~\bibnamefont {Nam}}, \bibinfo {author} {\bibfnamefont
  {E.}~\bibnamefont {Olsen}}, \bibinfo {author} {\bibfnamefont
  {J.}~\bibnamefont {Sarich}}, \ and\ \bibinfo {author} {\bibfnamefont
  {S.}~\bibnamefont {Wild}},\ }\href@noop {} {\bibfield  {journal} {\bibinfo
  {journal} {Comp. Phys. Comm.}\ }\textbf {\bibinfo {volume} {184}},\ \bibinfo
  {pages} {1592} (\bibinfo {year} {2013})}\BibitemShut {NoStop}%
\bibitem [{\citenamefont {Engel}\ \emph {et~al.}(1999)\citenamefont {Engel},
  \citenamefont {Bender}, \citenamefont {Dobaczewski}, \citenamefont
  {Nazarewicz},\ and\ \citenamefont {Surman}}]{Engel99}%
  \BibitemOpen
  \bibfield  {author} {\bibinfo {author} {\bibfnamefont {J.}~\bibnamefont
  {Engel}}, \bibinfo {author} {\bibfnamefont {M.}~\bibnamefont {Bender}},
  \bibinfo {author} {\bibfnamefont {J.}~\bibnamefont {Dobaczewski}}, \bibinfo
  {author} {\bibfnamefont {W.}~\bibnamefont {Nazarewicz}}, \ and\ \bibinfo
  {author} {\bibfnamefont {R.}~\bibnamefont {Surman}},\ }\href {\doibase
  10.1103/PhysRevC.60.014302} {\bibfield  {journal} {\bibinfo  {journal} {Phys.
  Rev. C}\ }\textbf {\bibinfo {volume} {60}},\ \bibinfo {pages} {014302}
  (\bibinfo {year} {1999})}\BibitemShut {NoStop}%
\bibitem [{\citenamefont {Reinhard}\ \emph {et~al.}(1999)\citenamefont
  {Reinhard}, \citenamefont {Dean}, \citenamefont {Nazarewicz}, \citenamefont
  {Dobaczewski}, \citenamefont {Maruhn},\ and\ \citenamefont
  {Strayer}}]{Reinhard1999}%
  \BibitemOpen
  \bibfield  {author} {\bibinfo {author} {\bibfnamefont {P.~G.}\ \bibnamefont
  {Reinhard}}, \bibinfo {author} {\bibfnamefont {D.~J.}\ \bibnamefont {Dean}},
  \bibinfo {author} {\bibfnamefont {W.}~\bibnamefont {Nazarewicz}}, \bibinfo
  {author} {\bibfnamefont {J.}~\bibnamefont {Dobaczewski}}, \bibinfo {author}
  {\bibfnamefont {J.~A.}\ \bibnamefont {Maruhn}}, \ and\ \bibinfo {author}
  {\bibfnamefont {M.~R.}\ \bibnamefont {Strayer}},\ }\href@noop {} {\bibfield
  {journal} {\bibinfo  {journal} {Phys. Rev. C}\ }\textbf {\bibinfo {volume}
  {60}},\ \bibinfo {pages} {014316} (\bibinfo {year} {1999})}\BibitemShut
  {NoStop}%
\bibitem [{\citenamefont {Schunck}\ \emph {et~al.}(2010)\citenamefont
  {Schunck}, \citenamefont {Dobaczewski}, \citenamefont {McDonnell},
  \citenamefont {Mor{\'e}}, \citenamefont {Nazarewicz}, \citenamefont
  {Sarich},\ and\ \citenamefont {Stoitsov}}]{Schunck2010}%
  \BibitemOpen
  \bibfield  {author} {\bibinfo {author} {\bibfnamefont {N.}~\bibnamefont
  {Schunck}}, \bibinfo {author} {\bibfnamefont {J.}~\bibnamefont
  {Dobaczewski}}, \bibinfo {author} {\bibfnamefont {J.}~\bibnamefont
  {McDonnell}}, \bibinfo {author} {\bibfnamefont {J.}~\bibnamefont {Mor{\'e}}},
  \bibinfo {author} {\bibfnamefont {W.}~\bibnamefont {Nazarewicz}}, \bibinfo
  {author} {\bibfnamefont {J.}~\bibnamefont {Sarich}}, \ and\ \bibinfo {author}
  {\bibfnamefont {M.~V.}\ \bibnamefont {Stoitsov}},\ }\href@noop {} {\bibfield
  {journal} {\bibinfo  {journal} {Phys. Rev. C}\ }\textbf {\bibinfo {volume}
  {81}},\ \bibinfo {pages} {024316} (\bibinfo {year} {2010})}\BibitemShut
  {NoStop}%
\bibitem [{\citenamefont {Munson}\ \emph {et~al.}(2012)\citenamefont {Munson},
  \citenamefont {Sarich}, \citenamefont {Wild}, \citenamefont {Benson},\ and\
  \citenamefont {McInnes}}]{Munson2012}%
  \BibitemOpen
  \bibfield  {author} {\bibinfo {author} {\bibfnamefont {T.}~\bibnamefont
  {Munson}}, \bibinfo {author} {\bibfnamefont {J.}~\bibnamefont {Sarich}},
  \bibinfo {author} {\bibfnamefont {S.}~\bibnamefont {Wild}}, \bibinfo {author}
  {\bibfnamefont {S.}~\bibnamefont {Benson}}, \ and\ \bibinfo {author}
  {\bibfnamefont {L.~C.}\ \bibnamefont {McInnes}},\ }\href@noop {} {\emph
  {\bibinfo {title} {TAO 2.0 Users Manual}}},\ \bibinfo {type} {Tech. Rep.}\
  \bibinfo {number} {ANL/MCS-TM-322}\ (\bibinfo  {institution} {Mathematics and
  Computer Science Division, Argonne National Laboratory},\ \bibinfo {year}
  {2012})\BibitemShut {NoStop}%
\bibitem [{\citenamefont {Towns}\ \emph {et~al.}(2014)\citenamefont {Towns},
  \citenamefont {Cockerill}, \citenamefont {Dahan}, \citenamefont {Foster},
  \citenamefont {Gaither}, \citenamefont {Grimshaw}, \citenamefont {Hazlewood},
  \citenamefont {Lathrop}, \citenamefont {Lifka}, \citenamefont {Peterson},
  \citenamefont {Roskies}, \citenamefont {Scott},\ and\ \citenamefont
  {Wilkins-Diehr}}]{Towns2014}%
  \BibitemOpen
  \bibfield  {author} {\bibinfo {author} {\bibfnamefont {J.}~\bibnamefont
  {Towns}}, \bibinfo {author} {\bibfnamefont {T.}~\bibnamefont {Cockerill}},
  \bibinfo {author} {\bibfnamefont {M.}~\bibnamefont {Dahan}}, \bibinfo
  {author} {\bibfnamefont {I.}~\bibnamefont {Foster}}, \bibinfo {author}
  {\bibfnamefont {K.}~\bibnamefont {Gaither}}, \bibinfo {author} {\bibfnamefont
  {A.}~\bibnamefont {Grimshaw}}, \bibinfo {author} {\bibfnamefont
  {V.}~\bibnamefont {Hazlewood}}, \bibinfo {author} {\bibfnamefont
  {S.}~\bibnamefont {Lathrop}}, \bibinfo {author} {\bibfnamefont
  {D.}~\bibnamefont {Lifka}}, \bibinfo {author} {\bibfnamefont {G.~D.}\
  \bibnamefont {Peterson}}, \bibinfo {author} {\bibfnamefont {R.}~\bibnamefont
  {Roskies}}, \bibinfo {author} {\bibfnamefont {J.~R.}\ \bibnamefont {Scott}},
  \ and\ \bibinfo {author} {\bibfnamefont {N.}~\bibnamefont {Wilkins-Diehr}},\
  }\href@noop {} {\bibfield  {journal} {\bibinfo  {journal} {Computing in
  Science \& Engineering}\ }\textbf {\bibinfo {volume} {16}},\ \bibinfo {pages}
  {62} (\bibinfo {year} {2014})}\BibitemShut {NoStop}%
\bibitem [{\citenamefont {Kortelainen}\ \emph {et~al.}(2010)\citenamefont
  {Kortelainen}, \citenamefont {Furnstahl}, \citenamefont {Nazarewicz},\ and\
  \citenamefont {Stoitsov}}]{Kortelainen2010}%
  \BibitemOpen
  \bibfield  {author} {\bibinfo {author} {\bibfnamefont {M.}~\bibnamefont
  {Kortelainen}}, \bibinfo {author} {\bibfnamefont {R.~J.}\ \bibnamefont
  {Furnstahl}}, \bibinfo {author} {\bibfnamefont {W.}~\bibnamefont
  {Nazarewicz}}, \ and\ \bibinfo {author} {\bibfnamefont {M.~V.}\ \bibnamefont
  {Stoitsov}},\ }\href@noop {} {\bibfield  {journal} {\bibinfo  {journal}
  {Phys. Rev. C}\ }\textbf {\bibinfo {volume} {82}},\ \bibinfo {pages}
  {011304(R)} (\bibinfo {year} {2010})}\BibitemShut {NoStop}%
\bibitem [{\citenamefont {Warburton}(1991)}]{Warburton1991}%
  \BibitemOpen
  \bibfield  {author} {\bibinfo {author} {\bibfnamefont {E.~K.}\ \bibnamefont
  {Warburton}},\ }\href@noop {} {\bibfield  {journal} {\bibinfo  {journal}
  {Phys. Rev. C}\ }\textbf {\bibinfo {volume} {44}},\ \bibinfo {pages} {233}
  (\bibinfo {year} {1991})}\BibitemShut {NoStop}%
\bibitem [{\citenamefont {Lorusso}\ \emph {et~al.}(2015)\citenamefont
  {Lorusso}, \citenamefont {Nishimura}, \citenamefont {Xu}, \citenamefont
  {Jungclaus}, \citenamefont {Shimizu}, \citenamefont {Simpson}, \citenamefont
  {S\"oderstr\"om}, \citenamefont {Watanabe}, \citenamefont {Browne},
  \citenamefont {Doornenbal}, \citenamefont {Gey}, \citenamefont {Jung},
  \citenamefont {Meyer}, \citenamefont {Sumikama}, \citenamefont {Taprogge},
  \citenamefont {Vajta}, \citenamefont {Wu}, \citenamefont {Baba},
  \citenamefont {Benzoni}, \citenamefont {Chae}, \citenamefont {Crespi},
  \citenamefont {Fukuda}, \citenamefont {Gernh\"auser}, \citenamefont {Inabe},
  \citenamefont {Isobe}, \citenamefont {Kajino}, \citenamefont {Kameda},
  \citenamefont {Kim}, \citenamefont {Kim}, \citenamefont {Kojouharov},
  \citenamefont {Kondev}, \citenamefont {Kubo}, \citenamefont {Kurz},
  \citenamefont {Kwon}, \citenamefont {Lane}, \citenamefont {Li}, \citenamefont
  {Montaner-Piz\'a}, \citenamefont {Moschner}, \citenamefont {Naqvi},
  \citenamefont {Niikura}, \citenamefont {Nishibata}, \citenamefont {Odahara},
  \citenamefont {Orlandi}, \citenamefont {Patel}, \citenamefont {Podoly\'ak},
  \citenamefont {Sakurai}, \citenamefont {Schaffner}, \citenamefont {Schury},
  \citenamefont {Shibagaki}, \citenamefont {Steiger}, \citenamefont {Suzuki},
  \citenamefont {Takeda}, \citenamefont {Wendt}, \citenamefont {Yagi},\ and\
  \citenamefont {Yoshinaga}}]{Lorusso2015}%
  \BibitemOpen
  \bibfield  {author} {\bibinfo {author} {\bibfnamefont {G.}~\bibnamefont
  {Lorusso}}, \bibinfo {author} {\bibfnamefont {S.}~\bibnamefont {Nishimura}},
  \bibinfo {author} {\bibfnamefont {Z.~Y.}\ \bibnamefont {Xu}}, \bibinfo
  {author} {\bibfnamefont {A.}~\bibnamefont {Jungclaus}}, \bibinfo {author}
  {\bibfnamefont {Y.}~\bibnamefont {Shimizu}}, \bibinfo {author} {\bibfnamefont
  {G.~S.}\ \bibnamefont {Simpson}}, \bibinfo {author} {\bibfnamefont {P.-A.}\
  \bibnamefont {S\"oderstr\"om}}, \bibinfo {author} {\bibfnamefont
  {H.}~\bibnamefont {Watanabe}}, \bibinfo {author} {\bibfnamefont
  {F.}~\bibnamefont {Browne}}, \bibinfo {author} {\bibfnamefont
  {P.}~\bibnamefont {Doornenbal}}, \bibinfo {author} {\bibfnamefont
  {G.}~\bibnamefont {Gey}}, \bibinfo {author} {\bibfnamefont {H.~S.}\
  \bibnamefont {Jung}}, \bibinfo {author} {\bibfnamefont {B.}~\bibnamefont
  {Meyer}}, \bibinfo {author} {\bibfnamefont {T.}~\bibnamefont {Sumikama}},
  \bibinfo {author} {\bibfnamefont {J.}~\bibnamefont {Taprogge}}, \bibinfo
  {author} {\bibfnamefont {Z.}~\bibnamefont {Vajta}}, \bibinfo {author}
  {\bibfnamefont {J.}~\bibnamefont {Wu}}, \bibinfo {author} {\bibfnamefont
  {H.}~\bibnamefont {Baba}}, \bibinfo {author} {\bibfnamefont {G.}~\bibnamefont
  {Benzoni}}, \bibinfo {author} {\bibfnamefont {K.~Y.}\ \bibnamefont {Chae}},
  \bibinfo {author} {\bibfnamefont {F.~C.~L.}\ \bibnamefont {Crespi}}, \bibinfo
  {author} {\bibfnamefont {N.}~\bibnamefont {Fukuda}}, \bibinfo {author}
  {\bibfnamefont {R.}~\bibnamefont {Gernh\"auser}}, \bibinfo {author}
  {\bibfnamefont {N.}~\bibnamefont {Inabe}}, \bibinfo {author} {\bibfnamefont
  {T.}~\bibnamefont {Isobe}}, \bibinfo {author} {\bibfnamefont
  {T.}~\bibnamefont {Kajino}}, \bibinfo {author} {\bibfnamefont
  {D.}~\bibnamefont {Kameda}}, \bibinfo {author} {\bibfnamefont {G.~D.}\
  \bibnamefont {Kim}}, \bibinfo {author} {\bibfnamefont {Y.-K.}\ \bibnamefont
  {Kim}}, \bibinfo {author} {\bibfnamefont {I.}~\bibnamefont {Kojouharov}},
  \bibinfo {author} {\bibfnamefont {F.~G.}\ \bibnamefont {Kondev}}, \bibinfo
  {author} {\bibfnamefont {T.}~\bibnamefont {Kubo}}, \bibinfo {author}
  {\bibfnamefont {N.}~\bibnamefont {Kurz}}, \bibinfo {author} {\bibfnamefont
  {Y.~K.}\ \bibnamefont {Kwon}}, \bibinfo {author} {\bibfnamefont {G.~J.}\
  \bibnamefont {Lane}}, \bibinfo {author} {\bibfnamefont {Z.}~\bibnamefont
  {Li}}, \bibinfo {author} {\bibfnamefont {A.}~\bibnamefont {Montaner-Piz\'a}},
  \bibinfo {author} {\bibfnamefont {K.}~\bibnamefont {Moschner}}, \bibinfo
  {author} {\bibfnamefont {F.}~\bibnamefont {Naqvi}}, \bibinfo {author}
  {\bibfnamefont {M.}~\bibnamefont {Niikura}}, \bibinfo {author} {\bibfnamefont
  {H.}~\bibnamefont {Nishibata}}, \bibinfo {author} {\bibfnamefont
  {A.}~\bibnamefont {Odahara}}, \bibinfo {author} {\bibfnamefont
  {R.}~\bibnamefont {Orlandi}}, \bibinfo {author} {\bibfnamefont
  {Z.}~\bibnamefont {Patel}}, \bibinfo {author} {\bibfnamefont
  {Z.}~\bibnamefont {Podoly\'ak}}, \bibinfo {author} {\bibfnamefont
  {H.}~\bibnamefont {Sakurai}}, \bibinfo {author} {\bibfnamefont
  {H.}~\bibnamefont {Schaffner}}, \bibinfo {author} {\bibfnamefont
  {P.}~\bibnamefont {Schury}}, \bibinfo {author} {\bibfnamefont
  {S.}~\bibnamefont {Shibagaki}}, \bibinfo {author} {\bibfnamefont
  {K.}~\bibnamefont {Steiger}}, \bibinfo {author} {\bibfnamefont
  {H.}~\bibnamefont {Suzuki}}, \bibinfo {author} {\bibfnamefont
  {H.}~\bibnamefont {Takeda}}, \bibinfo {author} {\bibfnamefont
  {A.}~\bibnamefont {Wendt}}, \bibinfo {author} {\bibfnamefont
  {A.}~\bibnamefont {Yagi}}, \ and\ \bibinfo {author} {\bibfnamefont
  {K.}~\bibnamefont {Yoshinaga}},\ }\href@noop {} {\bibfield  {journal}
  {\bibinfo  {journal} {Phys. Rev. Lett.}\ }\textbf {\bibinfo {volume} {114}},\
  \bibinfo {pages} {192501} (\bibinfo {year} {2015})}\BibitemShut {NoStop}%
\bibitem [{\citenamefont {Litvinova}\ \emph {et~al.}(2014)\citenamefont
  {Litvinova}, \citenamefont {Brown}, \citenamefont {Fang}, \citenamefont
  {Marketin},\ and\ \citenamefont {Zegers}}]{Litvinova2014}%
  \BibitemOpen
  \bibfield  {author} {\bibinfo {author} {\bibfnamefont {E.}~\bibnamefont
  {Litvinova}}, \bibinfo {author} {\bibfnamefont {B.~A.}\ \bibnamefont
  {Brown}}, \bibinfo {author} {\bibfnamefont {D.~L.}\ \bibnamefont {Fang}},
  \bibinfo {author} {\bibfnamefont {T.}~\bibnamefont {Marketin}}, \ and\
  \bibinfo {author} {\bibfnamefont {R.~G.~T.}\ \bibnamefont {Zegers}},\
  }\href@noop {} {\bibfield  {journal} {\bibinfo  {journal} {Phys. Lett. B}\
  }\textbf {\bibinfo {volume} {730}},\ \bibinfo {pages} {307} (\bibinfo {year}
  {2014})}\BibitemShut {NoStop}%
\bibitem [{\citenamefont {Shafer}()}]{Shafer}%
  \BibitemOpen
  \bibfield  {author} {\bibinfo {author} {\bibfnamefont {T.}~\bibnamefont
  {Shafer}},\ }\href@noop {} {\bibinfo  {journal} {in preparation}\
  }\BibitemShut {NoStop}%
\end{thebibliography}

%

\end{document}